\documentclass[12pt,preprint]{aastex}

\usepackage[version=3]{mhchem}

\shorttitle{Spectral Line Survey toward NGC 3627}
\shortauthors{Watanabe et al.}

\begin{document}


\title{A 3~mm Spectral Line Survey toward the Barred Spiral Galaxy NGC~3627}


\author{Yoshimasa~Watanabe\altaffilmark{1,2,3}, Yuri~Nishimura \altaffilmark{4,5}, Kazuo~Sorai\altaffilmark{6,1,2}, Nami~Sakai\altaffilmark{7}, Nario~Kuno\altaffilmark{1,2}}

\and

\author{Satoshi Yamamoto\altaffilmark{8, 9}}


\altaffiltext{1}{Division of Physics, Faculty of Pure and Applied Sciences, University of Tsukuba,  Tsukuba, Ibaraki 305-8571, Japan}
\altaffiltext{2}{Tomonaga Center for the History of the Universe, University of Tsukuba,  Tsukuba, Ibaraki 305-8571, Japan}
\altaffiltext{3}{College of Engineering, Nihon University, 1 Nakagawara, Tokusada, Tamuramachi, Koriyama, Fukushima 963-8642, Japan}
\altaffiltext{4}{Institute of Astronomy, School of Science, The University of Tokyo, 2-21-1 Osawa, Mitaka, Tokyo 181-0015, Japan}
\altaffiltext{5}{National Astronomical Observatory of Japan, 2-21-1, Osawa, Mitaka, Tokyo 181-8588, Japan}
\altaffiltext{6}{Department of Physics / Department of Cosmoscience, Hokkaido University, Kita 10, Nishi 8, Kita-ku, Sapporo, Hokkaido, 060-0810, Japan}
\altaffiltext{7}{RIKEN Cluster for Pioneering Research, 2-1, Hirosawa, Wako, Saitama 351-0198, Japan}
\altaffiltext{8}{Department of Physics, The University of Tokyo, 7-3-1 Hongo, Bunkyo-ku, Tokyo, 113-0033, Japan}
\altaffiltext{9}{Research Center for the Early Universe, The University of Tokyo, 7-3-1 Hongo, Bunkyo-ku, Tokyo, 113-0033, Japan}


\begin{abstract}
We conduct spectral line survey observations in the 3~mm band toward a spiral arm, a bar-end, and a nuclear region of the nearby barred spiral galaxy NGC~3627 with the IRAM 30~m telescope and the Nobeyama 45~m telescope.  Additional observations are performed toward the spiral arm and the bar-end in the 2~mm band.  We detect 8, 11, and 9 molecular species in the spiral arm, the bar-end, and the nuclear region, respectively.   Star-formation activities are different among the three regions, and in particular, the nucleus of NGC~3627 is known as a LINER/Seyfert~2 type nucleus.  In spite of these physical differences, the chemical composition shows impressive similarities among the three regions.  This result means that the characteristic chemical composition associated with these regions is insensitive to the local physical conditions such as star formation rate, because such local effects are smeared out by extended quiescent molecular gas on scales of 1 kpc.  Moreover, the observed chemical compositions are also found to be similar to those of molecular clouds in our Galaxy and the spiral arm of M51, whose elemental abundances are close to those in NGC~3627.  Therefore, this study provides us with a standard template of the chemical composition of extended molecular clouds with the solar metalicity in nearby galaxies.
\end{abstract}





\section{Introduction}
Various molecular species have been detected in external galaxies \citep{muller01,muller05} due to increasing observational sensitivity in radio astronomy.  To study the chemical composition of molecular gas without any preconception, a spectral line survey, scanning over a wide frequency range in an unbiased way, is a powerful technique.  Indeed, a number of spectral line survey observations have been carried out toward nuclear regions of external galaxies with single-dish radio telescopes \citep[e.g.,][]{Martin2006,Costagliola2011,Snell2011,Aladro2015,Nakajima2018} as well as radio interferometers \citep[e.g.,][]{Martin2011,Costagliola2015,Meier2015}.  These studies revealed contributions of intense UV field, X-ray radiation, and shocks to the chemical composition of molecular gas in nuclear regions.  In addition, \citet{Nishimura2016a, Nishimura2016b} conducted spectral line survey observations toward low-metalicity galaxies, the Large Magellanic Clouds and IC10, and found that their chemical compositions reflect the abundances of heavy elements.

Recently, chemical compositions at a molecular-cloud scale have been recognized to be important in order to connect star-formation studies in our Galaxy with those in external galaxies.  For this purpose, spectral line survey observations were conducted toward \textit{`quiescent'} molecular clouds in galactic disks other than nuclear regions in nearby galaxies \citep[e.g.,][]{Watanabe2014, Nishimura2016a}.  These observations revealed the chemical composition averaged over \textit{`quiescent'} molecular clouds without influences of nuclear activities.  In addition to the extragalactic works, wide-field mapping observations of various molecular lines have been performed toward giant molecular clouds in our Galaxy \citep{Kauffmann2017,Nishimura2017,Pety2017,Shimajiri2017,Watanabe2017}.  These observations showed that the $J=1-0$ lines of HCN and HCO$^+$ trace rather diffuse molecular gas ($n_{\rm H_2} \sim 10^{3}$~cm$^{-3}$), although these molecular lines have widely been used as a tracer of dense molecular gas in external galaxies.  On the other hand, \citet{Kauffmann2017} and \citet{Pety2017} revealed that N$_2$H$^+$ traces cold and dense molecular gas ($n_{\rm H_2} > 10^{3}$~cm$^{-3}$).  However, the chemical study of \textit{`quiescent'} molecular clouds in nearby galaxies with metal abundances similar to that of the solar vicinity has been reported only for M51 \citep{Watanabe2014}.  We have therefore conducted a spectral line survey toward the \textit{`quiescent'} molecular gas in the nearby galaxy NGC~3627.

\begin{figure}
\epsscale{0.99}
\plotone{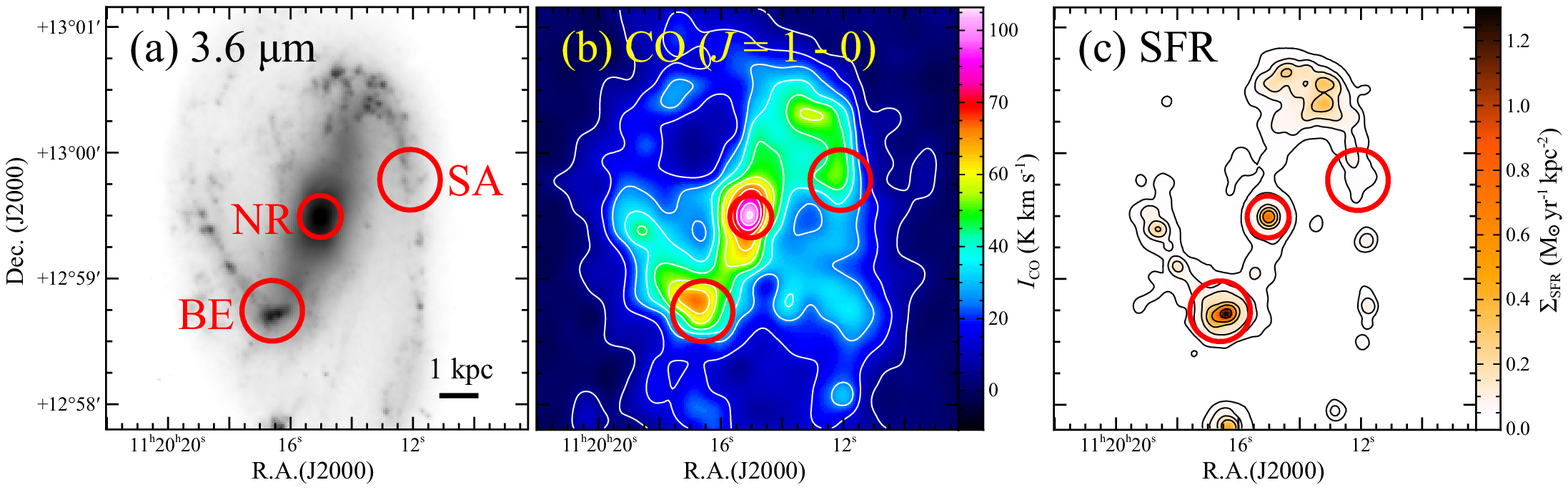}
\caption{Observed positions (red circles) in (a) the 3.6~$\mu$m image \citep{Kennicutt2003}, (b) the integrated intensity map of CO($J=1-0$) \citep{Kuno2007}, and (c) the surface density map of the star formation rate (SFR).  The contour levels are 5, 15, 25, ..., 105~K~km~s$^{-1}$ and 0.03, 0.06, 0.12, ..., 0.96~M$_{\odot}$~yr$^{-1}$~kpc$^{-2}$ for (b) and (c), respectively.  The observed positions are ($\alpha$ (J2000), $\delta$ (J2000)) = (11$^{\rm h}$ 20$^{\rm m}$ 12.2$^{\rm s}$, +12$^{\circ}$ 59$'$ 47.1$''$), (11$^{\rm h}$ 20$^{\rm m}$ 16.6$^{\rm s}$, +12$^{\circ}$ 58$'$ 44.6$''$), and (11$^{\rm h}$ 20$^{\rm m}$ 15.5$^{\rm s}$, +12$^{\circ}$ 59$'$ 29.6$''$), for NGC~3627~SA, NGC~3627~BE, and NGC~3627~NR, respectively.  The sizes of the red circles, which are $29''$ and $20''$, correspond to the largest observation beam sizes of the IRAM 30~m telescope and the Nobeyama~45~m telescope, respectively.  The star formation rate is derived from the H$\alpha$ and the 24~$\mu$m maps, which are observed by SINGS \citep{Kennicutt2003}, by using equation (3) \citep{Calzetti2007}.  The angular resolution of H$\alpha$ is convolved to $5''$, which is the same angular resolution of 24~$\mu$m data, before the deriving SFR. }
\label{fig01}
\end{figure}

NGC~3627 is a nearby barred spiral galaxy classified as SAB(s)b in the Third Reference Catalog of Bright Galaxies \citep[RC3:][]{deVaucouleurs1991} at a distance of 11~Mpc \citep{Saha1999}.  Distributions of molecular gas and star-formation activities have extensively been studied by CO observations with single-dish radio telescopes \citep[e.g.,][]{Kuno2007, Warren2010, Watanabe2011} and radio interferometers \citep[e.g.,][]{Helfer2003, Paladino2008, Casasola2011, Beuther2017,Gallagher2018a,Law2018,Sun2018}.  In the present study, we observe three regions of NGC~3627: a spiral arm, a bar-end, and a nuclear region (Table~\ref{tab01} and Figure~\ref{fig01}).  In the bar-end region, which connects a spiral arm and a bar, intense star formation activities are identified by the H$\alpha$ emission \citep{Smith1994,Sheth2002} and the free-free emission in the centimeter wavelength \citep{Paladino2008,Murphy2015}.  \citet{Watanabe2011} reported that the star formation rate (SFR) and the star formation efficiency (SFE) are both higher in the bar-ends than in the spiral arms and the nuclear region (Table~\ref{tab02}), where the SFRs are calculated from the H$\alpha$ and the 24~$\mu$m data \citep{Kennicutt2003} by using a calibration method given by \citet{Calzetti2007}.  With a high angular resolution observation of CO($J=2-1$), \citet{Beuther2017} found two velocity components of molecular gas in the bar-end which belong to the orbital families of the spiral arm and the bar.  They suggested that the high star formation activities are caused by interactions of molecular clouds at the interface between the spiral arm and the bar.  On the other hand, the nuclear region of NGC~3627 shows lower star formation activities than the bar-ends \citep{Regan2002,Watanabe2011,Murphy2015}, although it is known to possess the low-ionization nuclear emission region (LINER) or low-luminous Seyfert~2 \citep{Ho1997,Filho2000}.    

In this paper, we report the result of spectral line surveys toward the northern spiral arm (SA), the southern bar-end (BE), and the nuclear region (NR) of NGC~3627 (Figure~\ref{fig01}) with the IRAM 30~m and the Nobeyama 45~m radio telescopes.  By comparing the chemical composition among these three regions, we explore how chemical compositions at a scale of 1~kpc depend on galactic-scale physical environments.  

\section{Observations}
\subsection{IRAM 30~m Telescope}
Observations toward SA and BE in NG~3627 were performed with the IRAM~30~m telescope at Pico Veleta in July, 2014 and December, 2014.  The observed positions are shown in Figure~\ref{fig01} and Table~\ref{tab01}.  The beam sizes are 30--20$^{''}$ and 19--17$^{''}$ at the 3~mm and 2~mm band, respectively, which correspond to linear scales of 1.2--0.9~kpc and 0.8--0.7~kpc at the distance of 11.1~Mpc, respectively.  The observed frequency range is from 85.0 to 116.0~GHz and from 140.0 to 148.0~GHz for SA, while it is from 85.0 to 100.5~GHz, from 108.4 to 116.0~GHz, and from 140.0 to 148.0~GHz for BE.  The frequency range from 100.5 to 108.4~GHz does not involve strong spectral lines according to previous spectral line surveys toward the spiral arm of M51 \citep{Watanabe2014}.  Therefore, we put a lower priority on this frequency range, and did not observe it for BE because of the limited observation time.  Two EMIR (Eight MIxer Receivers) bands, E090 and E150, were used simultaneously with the dual polarization mode.  The EMIR is the sideband separating receiver, which outputs the USB and LSB signals separately.  The image rejection ratio is confirmed to be better than 13~dB and 10~dB for E090 and E150, respectively, according to the status report of the 30~m telescope\footnote{\url{http://www.iram.es/IRAMES/mainWiki/EmirforAstronomers}}.  The system noise temperature ranged from 75 to 310~K for E090 and from 100 to 210~K for E150.  Detailed frequency settings and system noise temperatures are summarized in Table \ref{tab03}.  Backends were eight FTS (Fourier Transform Spectrometers) autocorrelators whose bandwidth and channel width are 4050~MHz and 195~kHz, respectively.  The telescope pointing was checked every hour by observing the continuum source 1055+018 near the target position, and was found to be better than $\pm 5^{''}$.  The wobbler switching mode was employed with beam throw of $\pm120^{''}$ and switching frequency of 0.5~Hz.  The wobbler throw is toward the azimuth direction, and hence, the absolute off-position depends on the hour angle.  Nevertheless, the off-position was always out of the CO disk of NGC~3627.  The intensity scale was calibrated to the antenna temperature ($T_{\rm A}^{*}$) scale by using cold and hot loads.  $T_{\rm A}^{*}$ was converted to the main beam temperature $T_{\rm mb}$ by multiplying $F_{\rm eff}/B_{\rm eff}$.  Here, $F_{\rm eff}$ is the forward efficiency, which is 95~\% and 93~\% for the 3~mm and 2~mm bands, respectively, and $B_{\rm eff}$ is the main beam efficiency, which is 81~\% and 74~\% for the 3~mm and 2~mm bands, respectively.  Spectral baselines were subtracted for each correlator band (4050~MHz) and for each scan by fitting a 7th - 9th order polynomial to the line-free part before averaging all the scans.  Then, the final spectrum for each correlator band was obtained by integrating all the scans.  After the integration, the spectrum was smoothed to the frequency resolution of 5~MHz and 10~MHz for the 3~mm band and the 2~mm band, respectively (velocity resolution of $\sim 15$~km~s$^{-1}$ and $\sim 20$~km~s$^{-1}$, respectively).

\subsection{Nobeyama 45~m Telescope}
Observations toward the nuclear region of NGC~3627 (NGC~3627~NR: Figure~\ref{fig01} and Table~\ref{tab01}) were performed with the Nobeyama~45~m telescope in the 3~mm band in January and March, 2015.  The beam size is 20--15$^{''}$, which corresponds to a linear scale of 1.2--0.9~kpc at the distance of 11.1~Mpc.  The observed frequency range is from 85.0~GHz to 115.5~GHz.  The TZ1H/V receiver \citep{Nakajima2008} was used as the front end with the dual polarization mode.  It is the sideband separating receiver, which outputs the USB and LSB signals separately for the two linear polarization.  The image rejection ratio is assured to be better than 10~dB.  The system temperature ranged from 110 to 250~K.  Detailed frequency settings and system noise temperatures are summarized in Table \ref{tab03}.  The backends were 16 SAM45 autocorrelators \citep{Kamazaki2012} whose bandwidth and channel width each are $\sim 1600$~MHz and 0.5~MHz, respectively.  The telescope pointing was checked every hour by observing the SiO maser source W-Leo, and was {\rm confirmed} to be better than $\pm 6^{''}$.  The position switching mode was employed with an off-position {\rm separated} from the nuclear region by $10'$ along the azimuthal direction.  The intensity scale was calibrated to the antenna temperature ($T_{\rm A}^{*}$) scale by using the chopper-wheel method.  $T_{\rm A}^{*}$ was converted to the main beam temperature $T_{\rm mb}$ by dividing $B_{\rm eff}$.  Here, $B_{\rm eff}$ is the main beam efficiency, which is 41~\%, 37~\% and 30~\% at 86~GHZ, 110~GHz, and 115~GHz, respectively.  Spectral baselines were subtracted for each correlator band ($\sim 1600$~MHz) and for each scan by fitting a 6th - 9th order polynomial to the line-free part before averaging the scans.  Then, the final spectra for each correlator band were obtained by integrating all the scans.  After the integration, the spectrum was smoothed to the frequency resolution of 20~MHz corresponding to the velocity resolution of $\sim 60$~km~s$^{-1}$ in the 3~mm band.

\section{Results}
\subsection{Overall Spectra}
Figure~\ref{fig02} shows the composite spectra (Figures~\ref{fig02}a-c) and those with a magnified vertical scale (Figures~\ref{fig02}d-f) for NGC~3627~SA, NGC3627~BE, and NGC~3627~NR in the 3~mm band.  Typical rms noise levels of the 3~mm spectra are 0.4--1.3~mK, 0.5--1.6~mK, and 0.8--3.2~mK for SA, BE, and NR, respectively.  Figure~\ref{fig03} shows the expanded version of the 3~mm spectra.  In addition to the 3~mm band, the 2~mm band (140--148~GHz) was observed for SA and BE, as shown in Figure~\ref{fig04}.  Typical rms noise levels of the 2~mm spectra are 0.4~mK and 0.5--0.6~mK for SA and BE, respectively.  The frequency range, the frequency resolution, and the rms noise level for the overall spectra are summarized in Table~\ref{tab04}.  A line-of-sight velocity ($V_{\rm LSR}$) of 655~km~s$^{-1}$, 870~km~s$^{-1}$, and 715~km~s$^{-1}$ are used for SA, BE, and NR, respectively, in the spectra (Figures~\ref{fig02}, \ref{fig03}, and \ref{fig04}).  These velocities are derived from the $^{13}$CO($J=1-0$) mapping observation toward NGC~3627 with the Nobeyama 45~m telescope \citep{Watanabe2011}.

\subsection{Identified Molecules }
Tables~\ref{tab05}, \ref{tab06}, and \ref{tab07} summarize the parameters of emission lines detected in SA, BE, and NR, respectively.  We identify molecular emission lines with the aid of the spectral line databases, Cologne Database for Molecular Spectroscopy (CDMS) maintained by University of Cologne \citep{muller01,muller05} and Submillimeter, Millimeter and Microwave Spectral Line Catalog provided by Jet Propulsion Laboratory \citep{pickett98}.  A criterion for the line detection is that the peak intensity and the integrated intensity exceed $3\sigma$ evaluated from the rms noise at the expected frequency of the line.  If the peak intensity or the integrated intensity is less than $3\sigma$, we regarded the emission line as a marginal detection.  In addition to the detected molecular lines, the upper limits to the peak intensity and the integrated intensity for important undetected lines are also given in Tables~\ref{tab05}, \ref{tab06}, and \ref{tab07}.  $V_{\rm LSR}$ and the full width at half maxima (FWHM) of each emission line are derived by Gaussian fitting, except for CCH and CN.  $V_{\rm LSR}$ and FWHM are not derived for the two molecules, since the spectral lines are blended by several hyperfine structure components.  Although some lines show an asymmetric profile in BE, we simply applied Gaussian fitting to the spectra.   As the result, $V_{\rm LSR}$ is sometimes different from the velocity of intensity peak, and the uncertainties of $V_{\rm LSR}$ and FWHM are slightly larger than those in SA.  The derived $V_{\rm LSR}$ values are almost consistent with the $V_{\rm LSR}$ values obtained by the observations of $^{13}$CO($J=1-0$) \citep{Watanabe2011} in the three positions (Table~\ref{tab01}).  Moreover, we also confirm that the $V_{\rm LSR}$ values are consistent with those obtained by the other observations such as CO($J=1-0$) \citep{Sorai2019} and CO($J=2-1$) \citep{Leroy2009}. 

Figures~\ref{fig05}, \ref{fig06}, and \ref{fig07} display profiles of the lines listed in Tables~\ref{tab05}, \ref{tab06}, and \ref{tab07}.  In SA, 14 emission lines of 8 molecular species (CCH, CN, CO, HCN, HNC, HCO$^+$, H$_2$CO, and CS) and two isotopologues ($^{13}$CO and C$^{18}$O) are detected in the 3~mm and 2~mm bands.  In addition to these molecules, the emission line of CH$_3$OH is seen at 145.1~GHz with the signal-to-noise (S/N) ratio of 3 in the peak intensity and the integrated intensity.  However, we do not count CH$_3$OH in the detected molecule, because another emission line expected at 96.74~GHz is not detected.  In BE, 20 emission lines of 11 molecular species (CCH, CN, CO, HCN, HNC, HCO$^+$, H$_2$CO, CH$_3$OH, N$_2$H$^+$, CS, and SO) and 4 isotopologues ($^{13}$CO, C$^{18}$O, C$^{18}$O, and C$^{34}$S) are detected in the 3~mm and 2~mm bands.  In NR, 12 emission lines of 8 molecular species (CCH, CN, HCN, HNC, CO, HCO$^+$, CH$_3$OH, and CS) and two isotopologues ($^{13}$CO and C$^{18}$O) are detected in the 3~mm band.  Although one line of HC$_3$N ($J=11-10$; 100.076392~GHz) is detected with the S/N ratio of 3 and 4 in the peak intensity and the integrated intensity, respectively, no other transition lines, such as $J=10-9$ and $J=12-11$, is detected in NR.  Therefore, we treat HC$_3$N as a marginal detection.

The undetected lines in the Table~\ref{tab05}, \ref{tab06}, and \ref{tab07} include the lines of the isotopologues whose normal species lines are strongly detected, such as H$^{13}$CN, HN$^{13}$C, H$^{13}$CO$^+$, $^{13}$CS, and C$^{34}$S.  The tables also include the lines detected with relatively strong intensity in the nuclear region of other nearby galaxies such as c-C$_3$H$_2$, CH$_3$CCH, CH$_3$CN, HNCO, and OCS \citep[e.g.,][]{Aladro2013, Aladro2015, Lindberg2011, Martin2009, Nakajima2011, Nakajima2018}.  The hydrogen recombination lines of H42$\alpha$, H41$\alpha$, and H40$\alpha$ are also included as the undetected lines in Tables~\ref{tab05}, \ref{tab06}, and \ref{tab07}.  

%
%

\subsection{Column Densities}
Column densities of molecules are derived by assuming optically thin emission and local thermodynamic equilibrium (LTE) condition.  Since only one transition line is detected for each molecule except for BE, a rotation temperature cannot be derived from the observations.  We therefore adopt the three rotation temperatures, 10~K, 15~K, and 20~K, because a kinetic temperature of molecular gas is evaluated to be 10--20~K in the disk region of NGC~3627 \citep{Law2018}.  A temperature of cold dust in NGC~3627 is also found to fall in this range \citep{Galametz2012}.

Before deriving the column densities, the integrated intensities are divided by the beam filling factor $\theta^2_{\rm source}/(\theta^2_{\rm source}+\theta^2_{\rm beam})$, where $\theta_{\rm source}$ and $\theta_{\rm beam}$ are the source size and the FWHM of the telescope beam, respectively, to compensate for frequency dependence of the beam size.  Here, the source size ($\theta_{\rm source}$) is assumed to be $10''$, because distributions of HCO$^{+}(J=1-0)$ are resolved to be $\sim 10''$ with ALMA in BE and NR\citep{Murphy2015}.  We assume the same source size in SA for simplicity.  The $\theta_{\rm beam}$ value is $2460/f$~arcsec and $1710/f$~arcsec for the IRAM 30~m telescope and the Nobeyama 45~m telescope, respectively, where $f$ is the observation frequency in GHz.  

The column densities are evaluated by the following formula assuming the optically thin condition \citep[e.g.:][]{Goldsmith1999,Yamamoto2017}:
\begin{equation}
N = \frac{3W_{\nu}k_{\rm B} U(T_{\rm rot})}{8 \pi^3 S\mu_0^2 \nu} \exp\left(\frac{E_{\rm u}}{k_{\rm B}T_{\rm rot}}\right) \left\{ 1-\frac{\exp(h\nu/k_{\rm B}T_{\rm rot})-1}{\exp(h\nu/k_{\rm B}T_{\rm bg})-1}\right\}^{-1},
\label{eq01}
\end{equation}
where $W_{\nu}$ is integrated intensity, $k_{\rm B}$ is the Boltzmann constant, $U(T)$ is the partition function, $T_{\rm rot}$ is the assumed rotation temperature, $S$ is the line strength , $\mu_0$ is the dipole moment, $\nu$ is the frequency, $E_{\rm u}$ is the upper state energy, $h$ is the Planck constant, and $T_{\rm bg}$ is the cosmic microwave background temperature of 2.7~K.  The derived column densities are summarized in Tables~\ref{tab08}, \ref{tab09}, and \ref{tab10} for SA, BE, and NR, respectively.  The partition function $U(T)$ is numerically calculated by the following formula:
\begin{equation}
U(T) = \sum_i g_i \exp\left( -\frac{E_i}{k_{\rm B} T}\right),
\label{eq02}
\end{equation}
where $g_i$ and $E_i$ are the statical weight and the energy for the $i$-th state.  The energy levels up to 300~K or more listed in CDMS are involved in the calculation.  We confirm consistency between our values and those in CDMS at the temperature of 9.375~K and 18.75~K.

In addition to the column densities, the fractional abundances relative to the molecular hydrogen (H$_2$) are evaluated.  For this purpose, the column densities of H$_2$ are derived from the column densities of C$^{18}$O by assuming the [C$^{18}$O]/[H$_2$] ratio of $1.7 \times 10^{-7}$ \citep{Frerking1982,Goldsmith1997}.  Here, we assume that the [C$^{18}$O]/[H$_2$] ratio in NGC~3627 is the same as that in our Galaxy, because the elemental abundance in NGC~3627 is also similar to that in our Galaxy \citep{Contini2017}.  The derived fractional abundances are also summarized in Tables~\ref{tab08}, \ref{tab09}, and \ref{tab10} for SA, BE, and NR, respectively.  Figure~\ref{fig08} shows histograms of the fractional abundances derived at $T_{\rm rot}$ = 10~K for the three positions.

Variations of the fractional abundances by changing the assumed rotation temperature from 10~K to 20~K are smaller than those of the column densities for most molecules.  This is because the temperature effects are mostly compensated by taking the ratio of the two column densities.  In fact, variations of the fractional abundances are less than 10~\% except for CH$_3$OH, c-C$_3$H$_2$, CH$_3$CCH, HC$_3$N, and OCS, while those of the column densities are $10-30$~\%.  Although variation of the fractional abundances is slightly higher for CH$_3$OH, c-C$_3$H$_2$, and CH$_3$CCH, it is evaluated to be $\sim 20$~\% or less.  On the other hand, the fractional abundances of HC$_3$N and OCS show larger variation by a factor of two, because their column densities are derived by using the lines with relatively high upper state energies of $\sim 20$~K.  As shown in Equation \ref{eq01}, the column densities are proportional to $\exp(E_{\rm u}/k_{\rm B}T_{\rm rot})$, and hence, the column densities are sensitive to the assumed rotation temperature for the lines with {\boldmath $E_{\rm u}/k_{\rm B} > T_{\rm tot} $}.  However, this effect does not cause a serious impact on the following discussion, because only upper limits are derived for HC$_3$N and OCS.  In the following sections, we will use the fractional abundances relative to H$_2$ or C$^{18}$O to compare chemical compositions in different sources. 

\subsection{SFR and SFE}
The SFR (Figure~\ref{fig01}c) is calculated by using the H$\alpha$ and Spitzer 24~$\mu$m data \citep{Kennicutt2003}.  Before the calculation, H$\alpha$ image is convolved by a Gaussian kernel to be the angular resolution of $5''$ corresponding to the angular resolution of the 24~$\mu$m data.  The calibration of SFR is done by the following equation \citep{Calzetti2007}:
\begin{equation}
{\rm SFR} = 5.3 \times 10^{-42} [L({\rm H\alpha})_{\rm obs} + 0.031 \times L({\rm 24 \mu m})] \,({\rm M_{\odot} yr^{-1}}),
\label{eq02}
\end{equation}
where $L({\rm H\alpha})$ and $L({\rm 24 \mu m})$ are luminosities of the H$\alpha$ and 24~$\mu$m emission, respectively, in erg~s$^{-1}$.  The surface density of SFR ($\Sigma_{\rm SFR}$) is evaluated within the observation beam of $22''$, $22''$, and $16''$ for SA, BE, and NR, respectively.  The surface density of molecular gas mass $\Sigma_{\rm gas}$ is calculated from the column density of H$_2$ derived from the C$^{18}$O data assuming the rotation temperature of 10~K.  The contribution of He is involved in $\Sigma_{\rm gas}$ by multiplying a factor of 1.36.  Here, the source size of $10''$ is not taken into account for both $\Sigma_{\rm SFR}$ and $\Sigma_{\rm gas}$.  The SFE is calculated as $\Sigma_{\rm SFR}/\Sigma_{\rm gas}$.  The $\Sigma_{\rm SFR}$, $\Sigma_{\rm gas}$, and SFE values are summarized in Table~\ref{tab02}.

\subsection{Integrated Intensity Ratios of CO Isotopologues}
We calculate integrated intensity ratios of the CO isotopologues in the three positions.  The $^{12}$CO($J=1-0$)/$^{13}$CO($J=1-0$) intensity ratios are $11.9 \pm 0.3$, $10.8 \pm 0.2$, and $14 \pm 1$ in SA, BE, and NR, respectively.  These values are consistent with the intensity ratios in NGC~3627 previously observed with the Nobeyama 45~m telescope \citep{Watanabe2011} and the IRAM 30~m telescope \citep{Cormier2018}.  \citet{Law2018} reported the $^{12}$CO($J=2-1$)/$^{13}$CO($J=2-1$) ratios to be $\sim 4$ and $2.5 \pm 0.2$ in BE and NR, respectively, from the SMA observation at a resolution of $2.33'' \times 1.85''$.  Although the ratios are for the $J=2-1$ transition, these  values are much lower than our values.  As discussed in \citet{Law2018}, $^{12}$CO can trace much more diffuse molecular gas than that of $^{13}$CO and be detectable everywhere.  As the result, the ratios obtained with a larger beam size tend to be larger.  

However, this trend does not always work for any case.  For instance, \citet{Tan2011} observed the $^{12}$CO($J=1-0$)/$^{13}$CO($J=1-0$) ratio with the PMO 14~m telescope toward various positions in NGC~3627.  Although their ratio is consistent with our value in NR, their value of $5.3 \pm 1.3$ at the position covering BE is significantly lower than our value of $10.8 \pm 0.2$ at BE in spite of a larger observation beam ($60''$).  This implies the $^{12}$CO/$^{13}$CO ratio is sensitive to the physical properties and the distributions of molecular cloud involved in the beam.

The $^{13}$CO($J=1-0$)/C$^{18}$O($J=1-0$) intensity ratios observed in the present study are $6 \pm 1$, $5.4 \pm 0.6$, and $6 \pm 3$ in SA, BE, and NR, respectively.  The ratio for NR is similar to the mean intensity ratio of $6.0 \pm 0.9$ for the nuclear regions of nearby galaxies \citep{Jimenz2017}.  On the other hand, the ratios in SA and BE are found to be lower than the mean intensity ratio of $9.0 \pm 1.1$ in the disk region of NGC~3627 \citep{Jimenz2017}.

The lower ratios in SA and BE would originate from the optical depth effect.  If the optical depth of $^{13}$CO is higher in SA and BE than in the other disk regions, the ratio would be lower in SA and BE.  In our observation, the optical depth of the $^{13}$CO line are evaluated to be 0.018 for both SA and BE with the temperature of 15~K and the assumed source size of $10''$.  However, the optical depth of the $^{13}$CO line is likely higher than the above value, because distribution of molecular gas is expected to be clumpy.  In addition to the high angular resolution observation of $^{12}$CO in NGC~3627 with the radio interferometers \citep{Gallagher2018a,Gallagher2018b,Law2018,Sun2018}, observations of the CO isotopologues have recently been conducted with the ALMA at a scale of $1''$ \citep{Gallagher2018a}.  The result will be useful for examining the possibility of the clumpy structure in future.

Another possibility would be different $^{18}$O abundances for the different positions.  \citet{Jimenz2017} reported that the intensity ratios are lower in the vicinity of active star forming regions than quiescent regions among the nearby galaxies.  They claimed that the trend can be explained, for instance, by enhancement of the $^{18}$O abundance by nucleosynthesis in the massive stars. Since the SFR is relatively higher in SA and BE than the other disk regions in NGC~3627 (Table~\ref{tab02} and \citet{Watanabe2011}), the active star formation might result in the lower $^{13}$CO($J=1-0$)/C$^{18}$O($J=1-0$) intensity ratios in SA and BE in principle.  However, it is not obvious whether such an effect could cause the observable effect on the ratios.  Note that, for this reason, we employ the standard [C$^{18}$O]/[H$_2$] ratio to derive the H$_2$ column density, as mentioned in Section 3.3.  

\section{Discussion}
In this observation, we have systematically revealed chemical compositions in the spiral arm (SA), the bar-end (BE), and the nuclear region (NR) of NGC~3627 by the spectral line survey.  We here compare the fractional abundances among these three regions and discuss them in terms of physical environments and evolutionary stages in the following subsections.  In addition, we also compare the chemical composition in NGC~3627~BE with those in the spiral arm of nearby galaxy M51 and the molecular cloud W51 in our Galaxy.  Then, we compare the chemical composition in NR of NGC~3627 with that in the nuclear region of NGC~1068 to study effects of nuclear activities on the chemical composition.

\subsection{Comparisons of Chemical Compositions between SA and BE \label{SAandBE}}
Figure~\ref{fig09}a shows a correlation plot of the fractional abundances between SA and BE which are derived under the assumption of the rotation temperature of 10~K.  Although the SFE is higher in BE than in SA by a factor of 5 (Table~\ref{tab02}), the chemical composition is similar to each other.  In fact, the Pearson correlation coefficient is found to be 0.999.  Here, we do not take into account of the uncertainties in the calculation.  In addition, we exclude the upper limit values from the correlation-coefficient calculations.  The correlation coefficient changes by 0.001, even if the correlation coefficients are calculated from the fractional abundances assuming different rotation temperatures.  Because of the high SFE in BE, star formation feedbacks such as UV radiation from high-mass stars and shocks by outflows from protostars are expected to be prominent in BE, which would consequently affect the chemical composition in BE in comparison with those in SA.  However, no significant effect of such star formation feedbacks can be recognized in the chemical composition observed at a scale of $\sim 1$~kpc.  The feedback effects would be diluted by the large observation beam, while the chemical composition of extended molecular gas makes a dominant contribution to the spectrum.  Similar results have been reported in observations of nearby galaxies and Galactic molecular clouds \citep[e.g.,][]{Nishimura2016a, Nishimura2017, Watanabe2014, Watanabe2017}.

On the other hand, most of the fractional abundances are found to be systematically higher in BE than SA by a factors of 2-3, except for CCH and $^{13}$CO.  The difference cannot be explained by the assumed beam filling factors.  Even if we assume a smaller source size of $1''$ instead of $10''$, the fractional abundances become higher by only a factor of 1.1.  Therefore, the higher fractional abundances indicate that there is more dense molecular gas in BE than SA, because these molecular transition lines likely trace denser molecular gas than the gas traced by CO isotopologues due to their higher critical densities \citep[e.g.][]{Nishimura2017}.  

However, the relation of the higher fraction of dense gas to higher SFE in BE is not straightforward, because the relation itself is still controversial.  \citet{Muraoka2016} reported that volume density of H$_2$ ($n_{\rm H_2}$) is correlated with SFE in their observation of the nearby galaxy NGC~2903 with the Nobeyama 45~m telescope.  Since high $n_{\rm H_2}$ is expected in the region with the high dense gas fraction, the high dense gas fraction is a possible origin of the high SFE.  On the other hand, \citet{Law2018} found no correlation between the $n_{\rm H_2}$ and SFE in NGC~3627 at a scale of kpc.  In addition, \citet{Usero2015} also reported that no correlation can be seen between SFE and dense gas fraction evaluated from CO(1-0) and HCN(1-0) observed with IRAM 30~m telescope in nearby galaxies.  Therefore, we need more investigations on the relation in NGC~3627.

While the critical density of CCH is higher than those of CO isotopologues, the fractional abundance in SA is similar to that in BE.  This would be a result of the wider spatial distribution of CCH than those of the other molecules except for CO and its isotopologues.  In fact, the extended CCH distributions are found in the Galactic molecular clouds \citep[e.g.,][]{Nishimura2017, Pety2017}.  \citet{Nishimura2017} conducted a wide-field mapping observation of the Galactic molecular cloud W3(OH), and found that the CCH/HCO$^+$ ratio is higher in the outer regions of molecular clouds than the central parts.  According to \citet{Nishimura2017}, the abundance of CCH is high in the periphery of a molecular cloud where the UV photon can penetrate, because CCH is efficiently produced from C$^+$ ionized by the interstellar UV field.  Another reason would be the excitation conditions necessary for producing the CCH emission.  Because the dipole moment of CCH (0.77 Debye) is smaller than those of other molecules such as HCN (2.99 Debye) and CS (1.96 Debye), the critical density of CCH is lower than these molecules \citep[e.g.][]{Nishimura2017}.  Therefore, CCH can trace more diffuse gas than the other molecules due to chemical and physical reasons.

\subsection{Comparison with a Spiral Arm of M51 and a Galactic Molecular Cloud W51 \label{M51W51}}
Here, we compare the chemical composition in NGC~3627~BE with those in a spiral arm in M51 and a giant molecular cloud W51 in our Galaxy.  Since these sources are molecular clouds located in a disk region of galaxies, their chemical compositions are free from the effect of nuclear activities such as AGNs and starbursts.  The chemical composition in SA is similar to that in BE, and hence, we focus only on BE in this section.  

M51 is a nearby face-on spiral galaxy \citep[$d=8.4$~Mpc][]{Feldmeier1997,Vinko2012}.  The elemental abundance in M51 \citep[N/O:$\sim 0.25$, S/O:$0.03$:][]{Bresolin2004} is similar to that in NGC~3627 \citep[N/O:$\sim 0.23$, S/O:$\sim 0.025$:][]{Contini2017}.  Figure~\ref{fig09}b is a correlation plot of the molecular abundances between BE and M51~P1 \citep{Watanabe2014}. The rotation temperature assumed for BE is 10~K, and that for M51~P1 is 5~K.  The plot shows a good correlation with a correlation coefficient of 0.989.  The fractional abundances in BE are thus similar to those in M51~P1 for most of the species.  Because the beam size of the observation ($\sim 25''$) corresponds to $\sim 1$~kpc at the distance of M51, the spatial scale of the M51 observation is similar to that of our observation toward NGC~3627.  It is reported that the SFE is lower in M51~P1 ($5.9 \times 10^{-10}$~yr$^{-1}$) than in BE ($5.7 \times 10^{-9}$~yr$^{-1}$) by a factor of 10 \citep{Watanabe2014}.  In spite of the different SFE, the chemical compositions in BE and M51~P1 are similar to each other.  Therefore, the chemical compositions of molecular gas at a scale of $\sim 1$~kpc are thought to be unaffected by the star-formation activities.  

Figure~\ref{fig09}c shows a correlation plot between BE and the Galactic molecular cloud W51 \citep{Watanabe2017}. The rotation temperature assumed for BE is 10~K.  W51 is the molecular cloud complex with the vigorous massive star forming regions.  The fractional abundances in W51 were obtained from the spectrum stacked over the $39\,{\rm pc} \times 47\,{\rm pc}$ area centered at the W51A molecular cloud mapped with the Mopra 22~m telescope.  Therefore, the fractional abundances in W51 represent those at the scale of a molecular cloud.  The elemental abundance in the Galaxy is also similar to that in NGC~3627 \citep{Contini2017}.  A strong positive correlation (0.989) seen in Figure~\ref{fig09}c indicates that the chemical composition in BE is also similar to the molecular-cloud-scale ($\sim 50$~pc) chemical composition in the spiral arm of our Galaxy.

\subsection{Nuclear Region of NGC~3627}
Figures~\ref{fig09}d and \ref{fig09}e are correlation plots of the fractional abundances between SA and NR, and between BE and NR, where the rotation temperature of 10~K is assumed.  Although the fractional abundances are higher in NR than SA and BE by factors of $\sim 5$ and $\sim 2$, respectively, the correlation coefficients are as high as 0.991--0.993 for the both plots.  The strong positive correlation suggests that the chemical compositions are similar to each other between BE and NR, although NR possesses a LINER/Seyfert~2 nucleus \citep{Ho1997,Filho2000}.  The X-ray luminosity in the 2-10~keV band is reported to be $3.2 \times 10^{39}$~erg~s$^{-1}$, which is lower than that of prototypical Seyfert nuclei in NGC~1068 by three orders of magnitude \citep{Brightman2011}.   Therefore, the effect of nuclear activities would be limited to very small region ($< 100$~pc), and would not significantly affect the chemical composition in the major part of molecular gas in the beam. 

In spite of the overall similarity, we can see small differences between SA and NR and between BE and NR.  For example, the fractional abundances are higher in NR than SA and BE, except for CCH and N$_2$H$^+$.  The trend of higher fractional abundances in NR is not artificially caused by the assumption of the low rotation temperature of 10~K, although higher rotation temperatures were reported in the nuclear region of other galaxies at a scale of $\sim 1$~kpc \citep[e.g.][]{Aladro2011, Aladro2013, Nakajima2018}.  As seen in Table~\ref{tab10}, the fractional abundances derived by assuming the temperature of 20~K are similar to or slightly higher than those derived by assuming the temperature of 10~K.  The similar trend can be reproduced by assuming higher temperature (e.g. 100~K), except for HC$_3$N.  Therefore, even if the rotation temperature in NR is higher than 10~K due to the nuclear activity, we will see the same trend of higher fractional abundances in NR.  The higher fractional abundances would be a result of a higher fraction of the denser molecular gas in NR than in the other positions.  On the other hand, the abundances of CCH are similar among the three regions (Figure~\ref{fig09}d and \ref{fig09}e).  As discussed in Section~\ref{SAandBE}, CCH traces widely extended molecular gas as CO and its isotopologues rather than dense gas traced by other molecules.  Hence, the trend of CCH in Figures~\ref{fig09}d and \ref{fig09}e seems reasonable.  
 
The N$_2$H$^+$ abundance in NR is found to be similar to or lower than that in BE (Figure~\ref{fig09}e).  This result indicates inefficient production of N$_2$H$^+$ and/or efficient destruction working in NR.   There are two major destruction processes for N$_2$H$^+$; the reaction with CO and the dissociative recombination with an electron.  However the destruction by CO would not be the cause of the different N$_2$H$^+$ abundances, because the fractional abundances of various molecules, for which H$_2$ column density is evaluated from C$^{18}$O, are higher in NR than in SA and BE.  If the CO fractional abundance were enhanced in NR, the fractional abundances of the other molecules would be calculated to be lower in NR than in SA and BE.  Moreover, the dissociative recombination would not be the case, either.  Because the SFE is derived to be higher in BE than in NR, the UV radiation field is expected to be higher in BE than NR.  The electron abundance is consequently expected to be higher in BE than NR, which contradicts with the lower N$_2$H$^+$ abundance in NR.  However, the electron abundance could be enhanced by strong X-ray radiation in NR due to nuclear activities, which may contribute to destruction of N$_2$H$^+$ in NR.

\citet{Pety2017} reported that N$_2$H$^+$ traces the densest regions of Orion~B cloud, while HCN and HCO$^+$ traces the less dense molecular gas with H$_2$ density of $\sim 1500$~cm$^{-3}$.  They discussed that N$_2$H$^+$ can only survive in cold and dense gas where CO is frozen on the dust grain.  \citet{Kauffmann2017} also reported that N$_2$H$^+$(1-0) traces the dense gas at a characteristic H$_2$ density of $\sim 4000$~cm$^{-3}$, while HCN(1-0) can trace rather diffuse gas at the H$_2$ density of $\sim 870$~cm$^{-3}$, on the basis of the observations of the Orion molecular cloud.  If this picture is applied to the present case, a fraction of `cold dense gas' is lower in NR than in BE.  In any case, the lower abundance of N$_2$H$^+$ in NR seems to suggest some important physical conditions.  Interferometric studies revealing the distributions of N$_2$H$^+$ and other molecules are awaited for further discussions.

\subsection{Comparisons of the Nuclear Region with NGC~1068}
Here, we compare the chemical composition of NR with that of Seyfert~2 type AGN NGC~1068, because NGC~3627 possesses LINER or low-luminous Seyfert~2 AGN \citep{Ho1997,Filho2000}.  The AGNs are thought to be powered by accretion of material to the super-massive black hole at the center of galaxy and to emit strong X-ray.  As the result, the chemical compositions of molecular gas around the AGNs are expected to be affected by it (X-ray dominated region) \citep[e.g.,][]{Lepp1996,Maloney1996,Meijerink2005, Meijerink2007}.  The AGN activity is thought to be higher in NGC~1068 than in NGC~3627, because the intrinsic luminosity of X-ray in the 2-10~keV band is reported to be higher in NGC~1068 ($1.6 \times 10^{42}$~erg~s$^{-1}$) than in NGC~3627 ($3.2 \times 10^{39}$~erg~s$^{-1}$) by three orders of magnitude \citep{Brightman2011}.  

Figure~\ref{fig09}f shows a correlation plot of fractional abundances relative to C$^{18}$O between NR and the nuclear region of NGC~1068 \citep{Aladro2013,Aladro2015}.   Here, the rotation temperature assumed for NR is 10~K.  \citet{Aladro2015} derived column densities of molecules in NGC~1068 using the spectral line survey data observed with the IRAM 30~m telescope by assuming the rotation temperature of 10~K.  The fractional abundances of NGC~1068 are therefore the ones averaged over $\sim 1.5$~kpc.  The correlation coefficient of the plot is as high as 0.84.  Although it is slightly lower than those among the disk regions in nearby galaxies and our Galaxy (W51), it indicates that the chemical compositions are similar to each other despite different X-ray luminosities.  The similarity implies that the effect of X-ray is diluted by an overwhelming contribution of ambient molecular clouds due to the large observation beam ($\sim 1$~kpc).      

The fractional abundances are generally found to be higher in NGC~1068 than NGC~3627~NR by a factor of two except for CH$_3$OH, HC$_3$N, and CS, although the chemical composition are almost similar to each other as a whole.  The higher abundances would indicate a higher fraction of dense molecular gas in NGC~1068 than NGC~3627 NR.  On the other hand, the abundances of CH$_3$OH, CS, and HC$_3$N are higher in NGC~3627 NR than NGC~1068 by a factor of two, although HC$_3$N is tentative detection.  It is reported that CH$_3$OH and CS are enhanced by liberation from ice mantle of dust grains due to shocks in the Galactic sources, for instance, outflow shocks \citep[e.g.,][]{Bachiller1997}.  In external galaxies, the enhancement of CH$_3$OH is reported in the bar region, where shocks due to cloud-cloud collisions are expected \citep{Meier2005}.  It has also been seen in shocked regions caused by interactions of molecular gas in merging galaxies \citep{Saito2017,Ueda2017}.  However,  no sign of shocks such as outflows from the nuclear region and active star formation is reported for NGC~3627~NR even by high angular resolution observations with interferometers \citep{Casasola2011,Murphy2015,Law2018}.  Therefore, the shock would not be the case for the enhancement, although there may exist accretion shocks in the nuclear region of NGC~3627 in principle.  As for another possibility, one may think that these molecules could more efficiently be destroyed in NGC~1068 by strong radiation from the AGN.  However, CS, CH$_3$OH, and HC$_3$N are found to be associated with the circumnuclear disk in NGC~1068 by the ALMA observation \citep{Takano2014}.  Thus, the destruction by the radiation would not explain the differences, either.  An investigation of overabundant CH$_3$OH and CS in NGC~3627 will be an interesting future problem.

\subsection{Chemical Compositions of Molecular Clouds}
The chemical compositions of molecular clouds at a scale from 1~kpc to 10~pc have been studied for NGC~3627 and M51 \citep{Watanabe2014, Watanabe2016}, as well as the Galactic molecular clouds W51 \citep{Watanabe2017} and W3 \citep{Nishimura2017}.  All these sources have the similar elemental abundances.  From these studies, we find that the chemical composition of molecular gas is similar among different galaxies and among different regions with different star-formation/nuclear activities, when we observe them in the 3~mm band at a scale larger than a molecular cloud ($> 10$~pc).  Figures~\ref{fig10}a-e show the spectra observed in NGC~3627~SA, BE, NR, M51~P1, and W51 in the 3~mm bands.  Indeed, the relative intensities of molecular lines look similar among the five sources.  

On the other hand, the spectrum of the hot core (Orion~KL: Figure~\ref{fig09}f) observed at a scale of 0.04~pc is much different from the other spectra.  The difference reflects a different chemical composition in the hot core from those in molecular clouds as well as the different excitation condition; namely hotter and denser gas in the hot core than in molecular clouds.  At a scale of a molecular cloud core ($\sim 0.01$~pc), the chemical composition is largely controlled by its evolutionary state \citep[e.g.][]{Suzuki1992} and ambient UV environments \citep[e.g.,][]{Tielens1985}.  Therefore, we can make use of the chemical composition as tracers of their evolutionary stages and physical environments in the Galactic objects.  However, such chemical features of molecular cloud cores are smeared out, when we observe them at a scale larger than the molecular clouds.  The sound crossing time scale of the molecular clouds ($> 10^7$~years), which is estimated by assuming a typical size of 10~pc \citep[e.g.:][]{Roman-Duval2010} and a sound speed of 0.3~km~s$^{-1}$, is longer than the time scale of chemical equilibrium ($\sim 10^{5-6}$~years).  Here, the sound speed $c_{\rm s}$ is calculated by the following formula:
\begin{equation}
c_{\rm s} = \sqrt{\gamma \frac{k_{\rm B}}{\mu m_{\rm H}}T},
\label{eq03}
\end{equation}
where $\gamma$ is the specific heat ratio of gas (5/3), $\mu$ is the molecular mass, and $m_{\rm H}$ is the hydrogen mass.  We evaluated the sound speed with $\mu = 2$ and $T = 10$~K.  Therefore, the chemical composition in the molecular clouds is thought to be close to that in the chemical equilibrium state, which is almost independent of evolutionary history of molecular clouds.  

However, \citet{Harada2019} recently reported that the chemical compositions of the molecular clouds can be reproduced by their time-dependent chemical model with a relatively short time of $\sim 10^5$ years.  Because this time scale is much shorter than the lifetime of molecular cloud and the time scale for chemical equilibrium, they suggested that a turbulence in the molecular clouds regulates the chemical clock.  Namely, the chemical composition is refreshed by the interstellar UV radiation when the turbulence brings the molecular gas from the inside to the surface of the molecular cloud where the visual extinction is low.   

In spite of the similarity of the chemical composition at a scale of molecular clouds, a different chemical composition has been found in metal poor galaxies \citep{Nishimura2016a,Nishimura2016b}, shocked regions of merging galaxies \citep[e.g.][]{Saito2017,Ueda2017}, and nuclear regions of galaxies observed at a scale of a few hundred pc \citep[e.g.,][]{Costagliola2015,Meier2015,Ando2017}.  In these objects, the chemical compositions are thought to reflect low elemental abundances, gas dynamics at molecular cloud scales, and extreme environments due to feedback from nuclear activity and vigorous star-formation.  To highlight chemical characteristics in these objects, the result of this study can be used as a `standard' template of chemical composition without such influences.


\acknowledgments
The authors are grateful to the IRAM 30~m telescope staff and the Nobeyama Radio Observatory (NRO) staff for excellent support in the observation runs.  This work is based on observations carried out under project numbers 089-14 and 203-14 with the IRAM 30~m telescope. IRAM is supported by INSU/CNRS (France), MPG (Germany) and IGN (Spain).  The 45~m radio telescope is operated by the NRO, a branch of the National Astronomical Observatory of Japan, National Institutes of Natural Sciences.  This study is supported by a Grant-in-Aid from the Ministry of Education, Culture, Sports, Science, and Technology of Japan (No. 25108005, 18H05222, and 16K17657).



{\it Facilities:} \facility{IRAM~30~m, Nobeyama~45~m}.


\clearpage


\begin{figure}
\epsscale{0.89}
\plotone{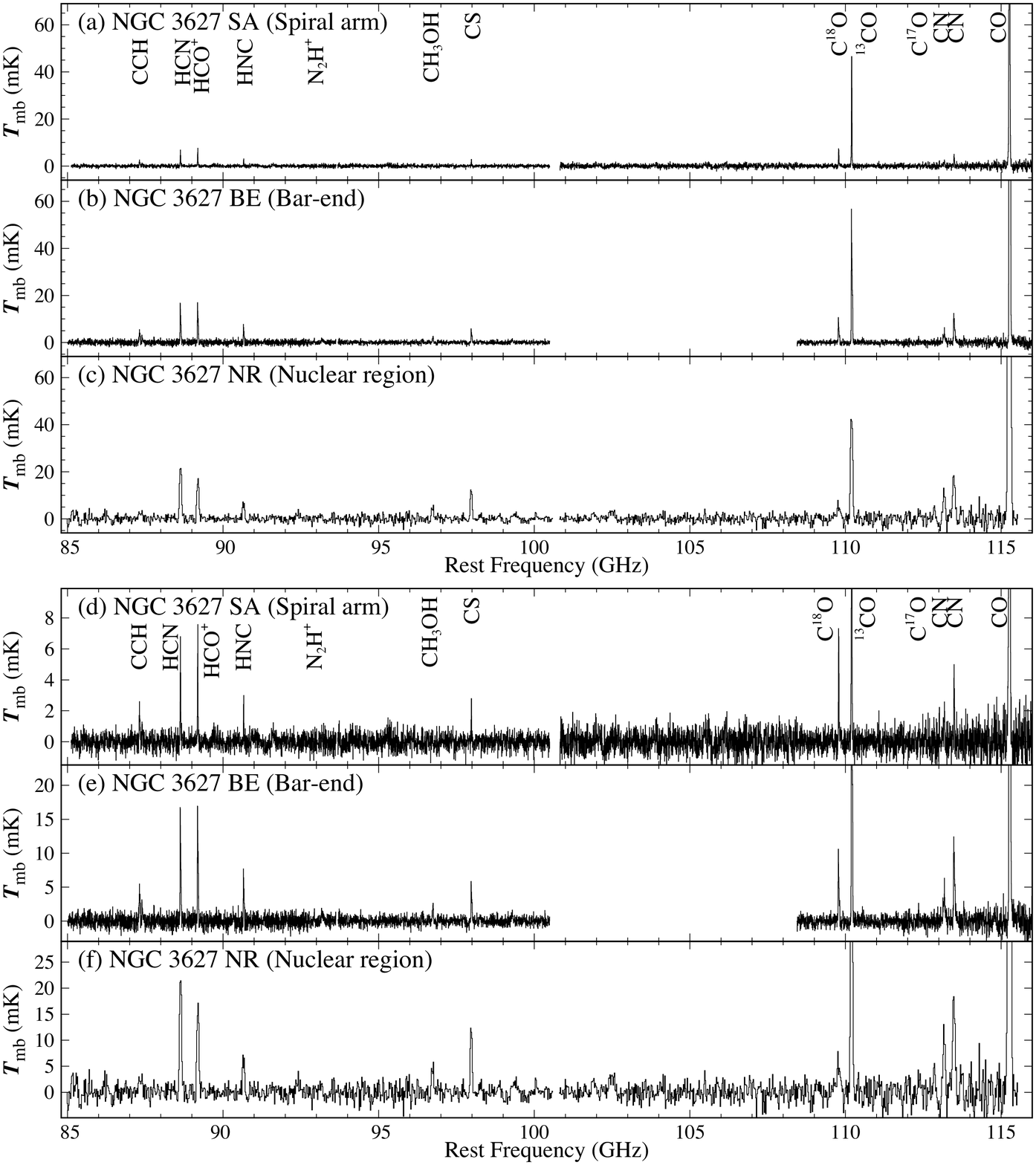}
\caption{Composite spectra of (a) NGC~3627~SA, (b) NGC~3627~BE, and (c) NGC~3627~NR in the 3~mm band and those with magnified vertical scales of (d) NGC~3627~SA, (e) NGC~3627~BE, and (f) NGC~3627~NR.  The observed positions are given in Table~\ref{tab01}. The frequency resolution is 5~MHz, and 20~MHz, for SA and BE, and NR, respectively.  The $V_{\rm LSR}$ values of 655~km~s$^{-1}$, 870~km~s$^{-1}$, and 715~km~s$^{-1}$ are assumed for SA, BE, and NR, respectively.  The emission lines contaminated from the image side band are removed.}
\label{fig02}
\end{figure}

\begin{figure}
\epsscale{0.99}
\plotone{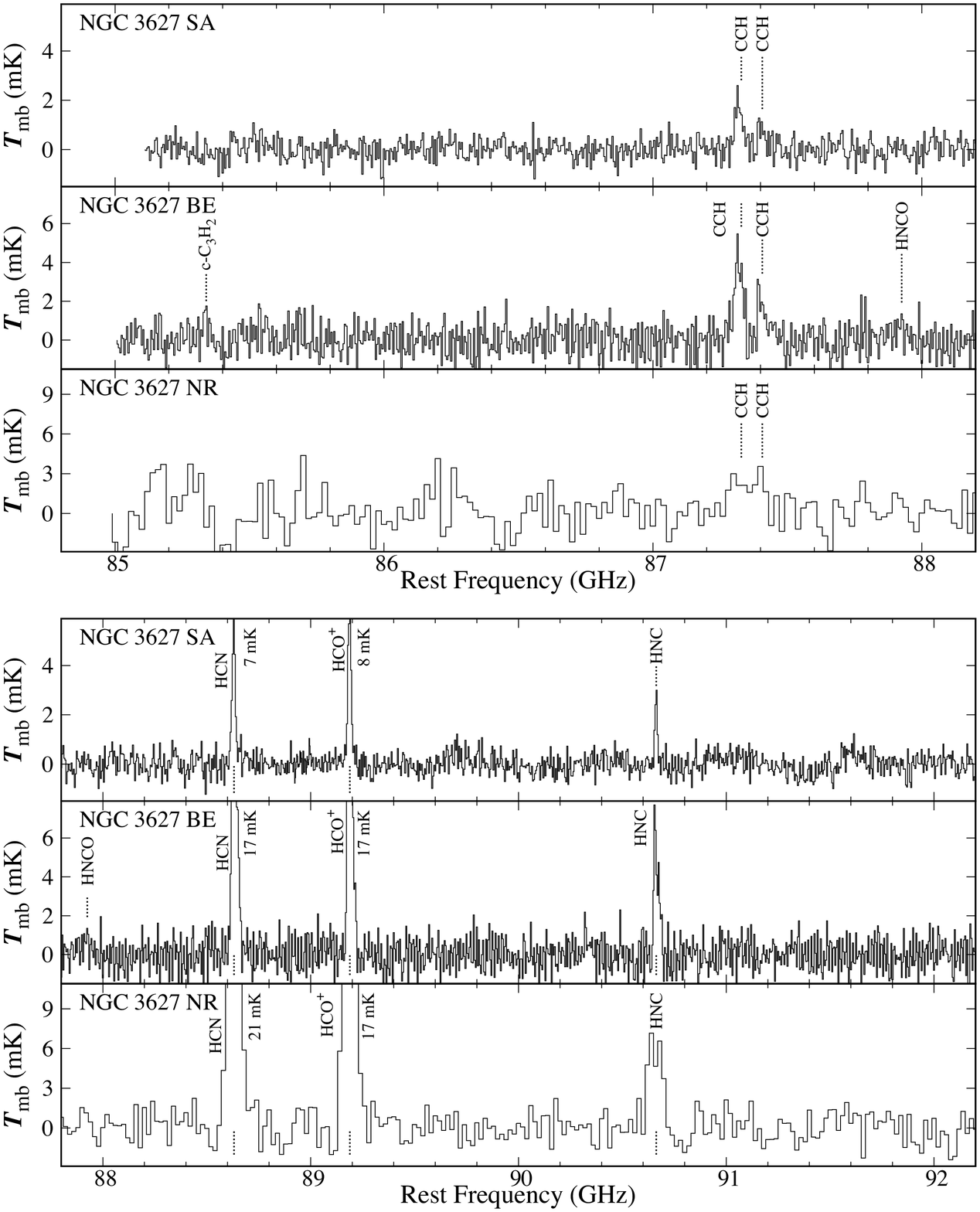}
\caption{Spectra of NGC~3627~SA, NGC~3617~BE, and NGC~3627~NR in the 3~mm band.  The spectra show the expanded version of Figure~\ref{fig02}.   The velocity resolutions and the assumed system velocities are the same as those in Figure~\ref{fig02}.  The mark `i-' indicates a contamination spectrum from the image side band. }
\label{fig03}
\addtocounter{figure}{-1}
\end{figure}

\begin{figure}
\epsscale{0.99}
\plotone{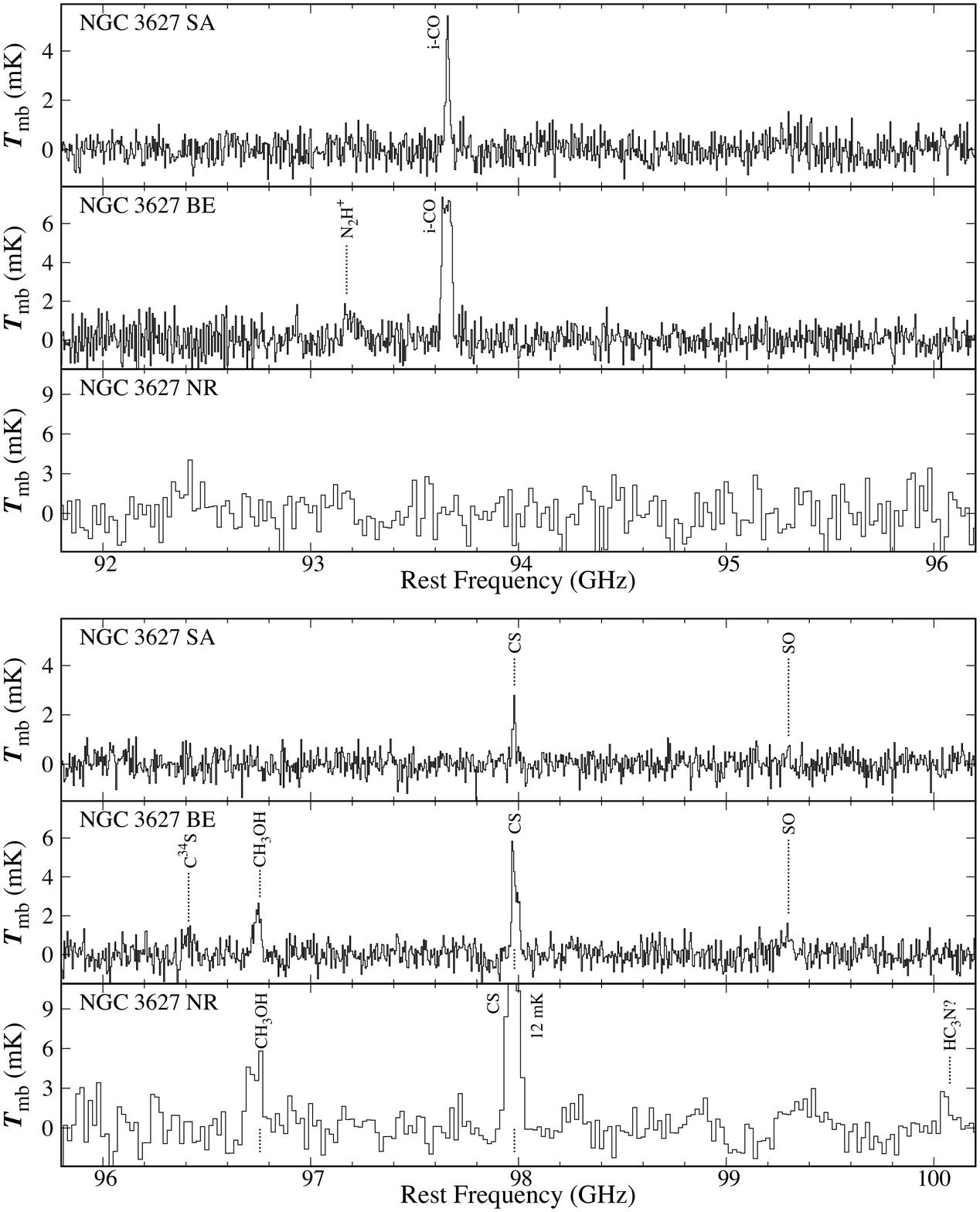}
\caption{(\textit{Continued})}
\addtocounter{figure}{-1}
\end{figure}

\begin{figure}
\epsscale{0.99}
\plotone{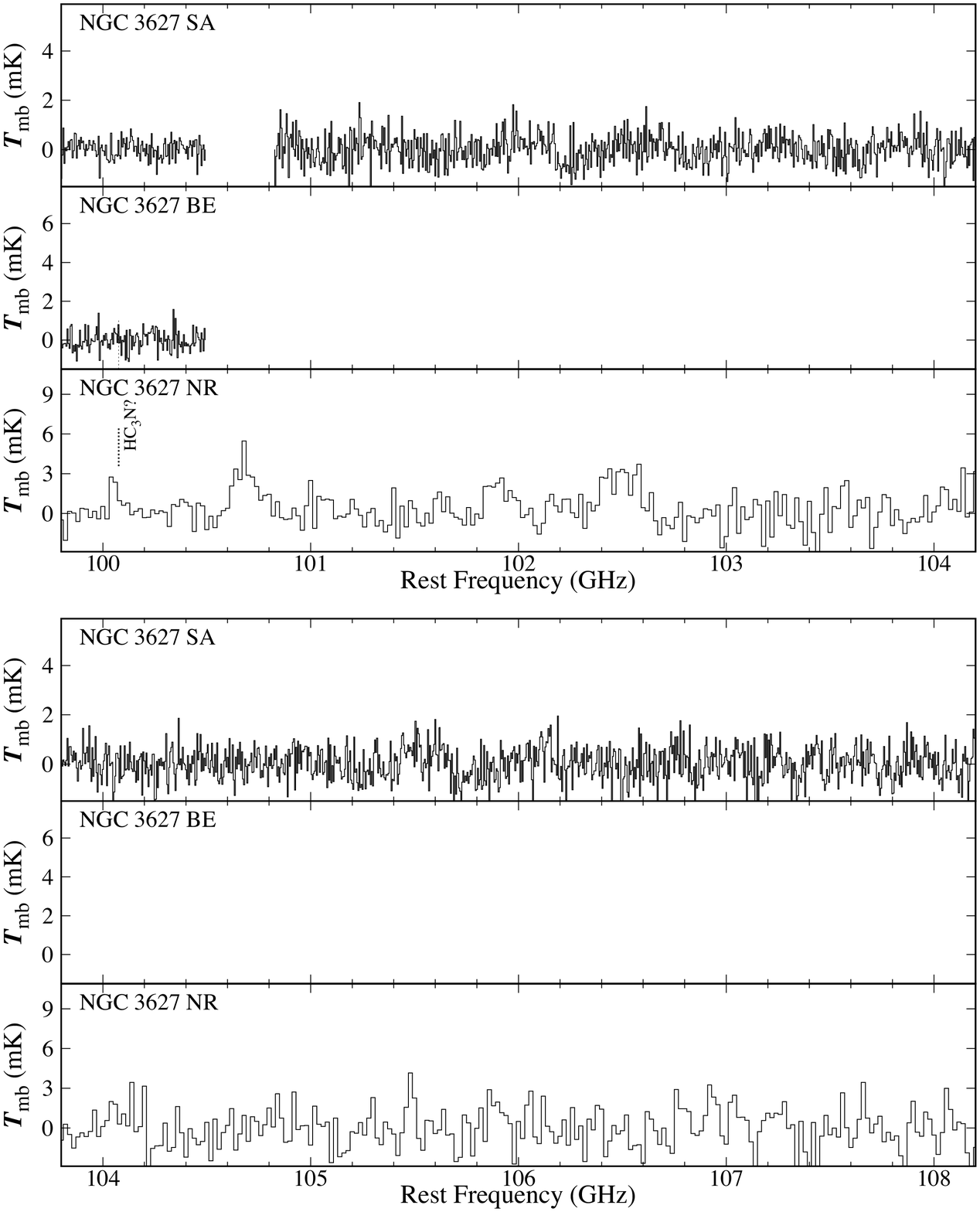}
\caption{(\textit{Continued})}
\addtocounter{figure}{-1}
\end{figure}

\begin{figure}
\epsscale{0.99}
\plotone{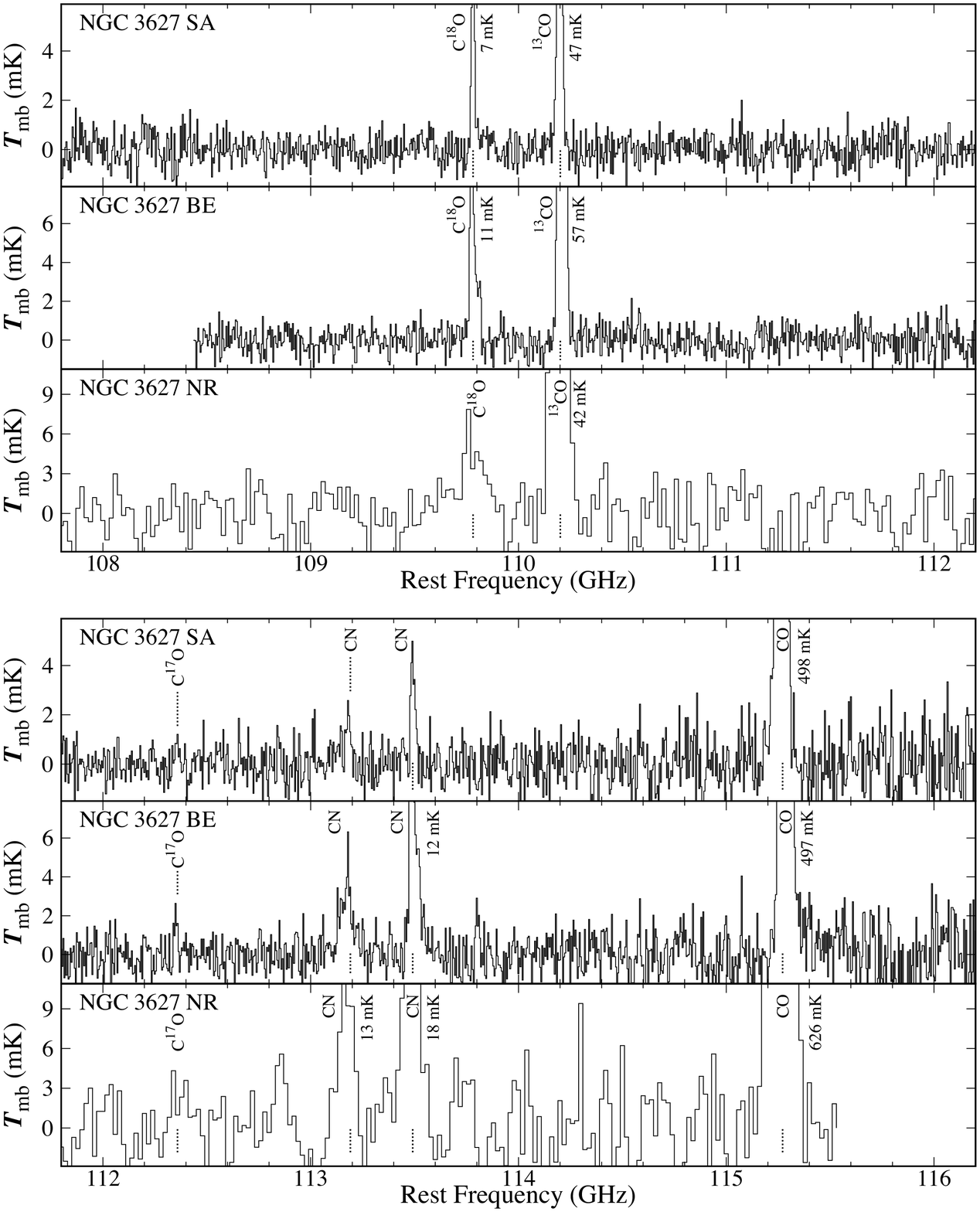}
\caption{(\textit{Continued})}
\end{figure}

\begin{figure}
\epsscale{0.99}
\plotone{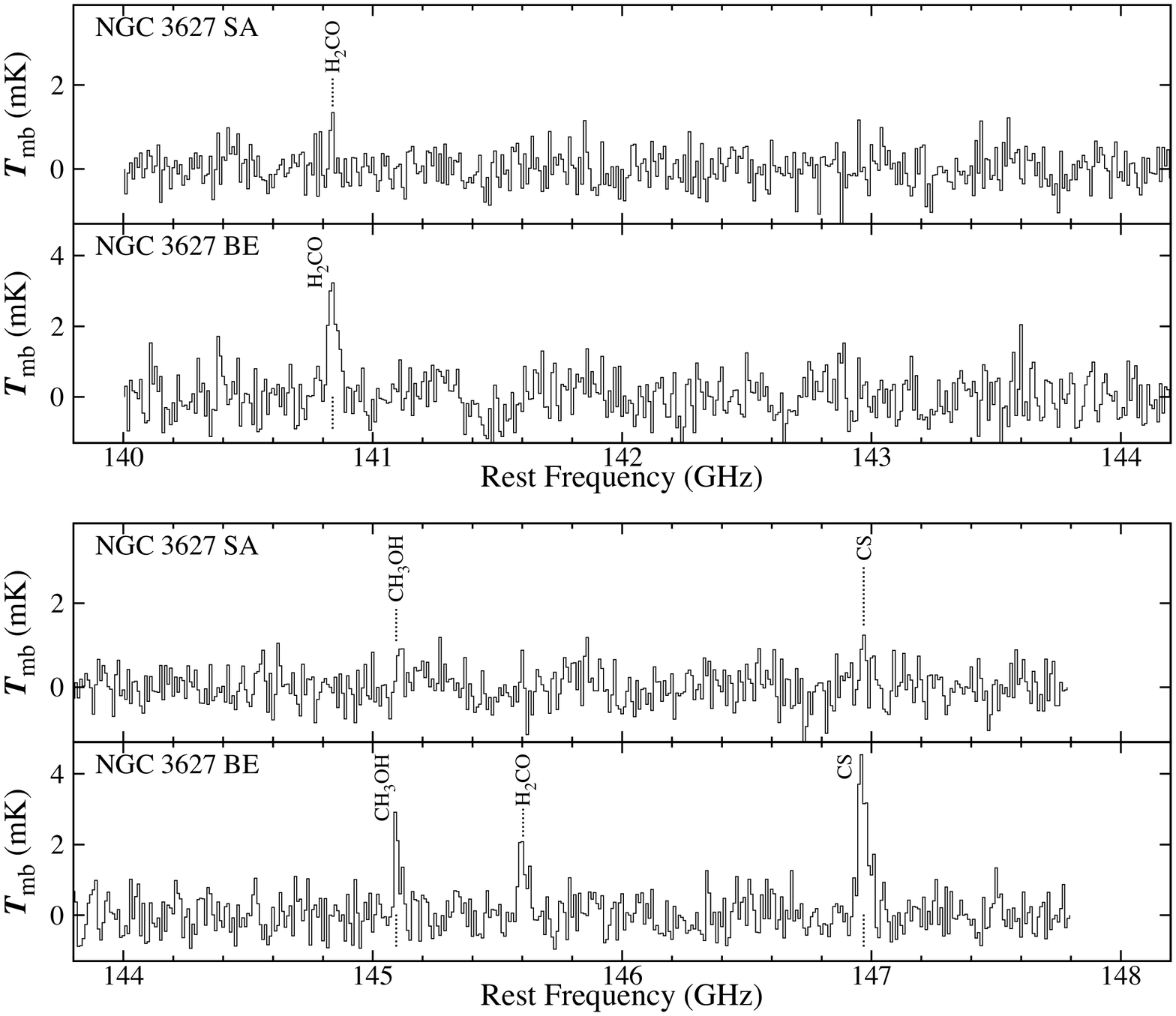}
\caption{Spectra of (a) NGC~3627~SA and (b) NGC~3627~BE in the 2~mm band. The frequency resolution is 10~MHz.  The $V_{\rm LSR}$ values of 655~km~s$^{-1}$ and 870~km~s$^{-1}$ are assumed for SA and BE, respectively. }
\label{fig04}
\end{figure}

\begin{figure}
\epsscale{0.99}
\plotone{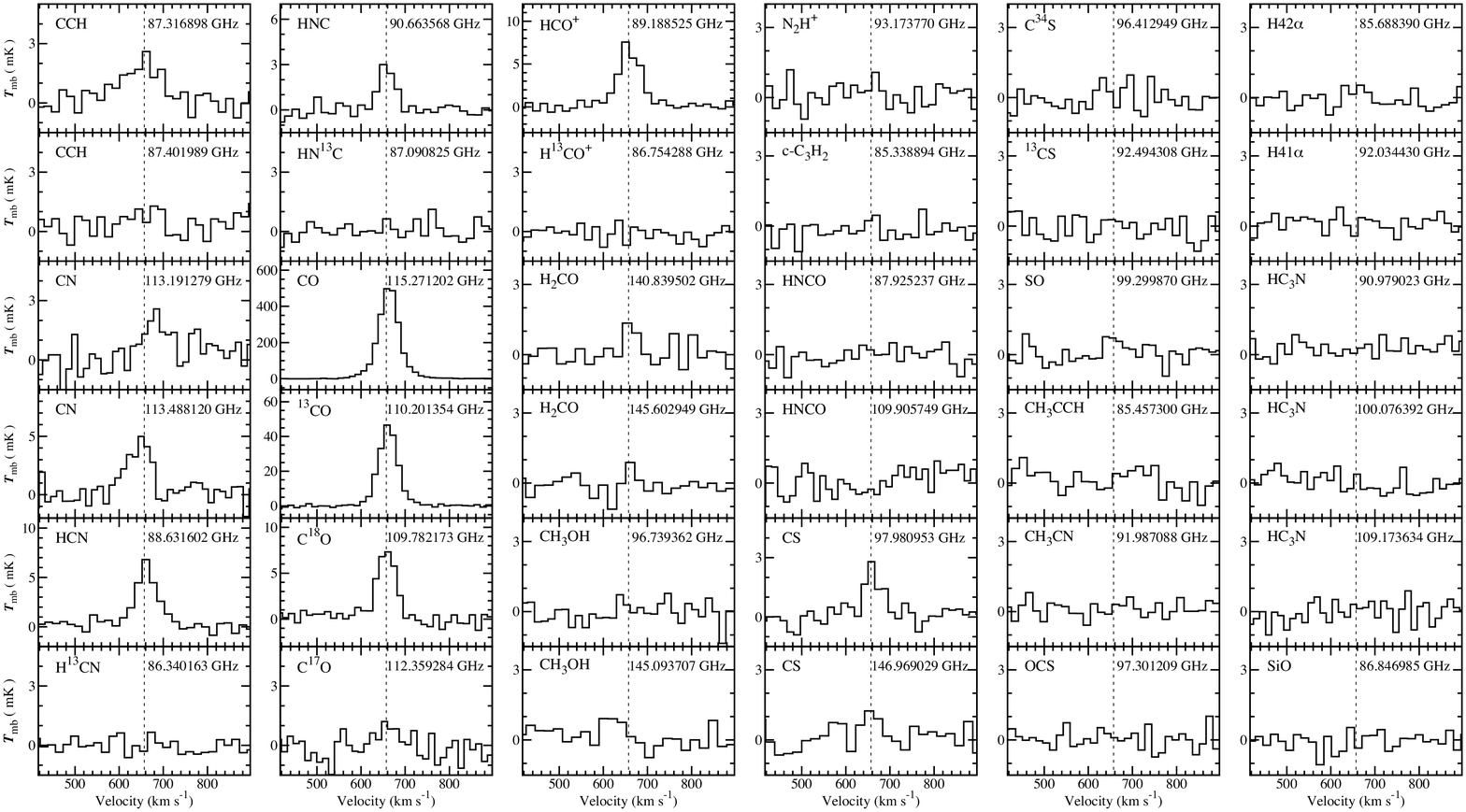}
\caption{Line profile of each molecular transition observed in SA.  Vertical dashed lines indicate $V_{\rm LSR}$ of 657~km~s$^{-1}$, which is the typical systemic velocity of SA. }
\label{fig05}
\end{figure}

\begin{figure}
\epsscale{0.99}
\plotone{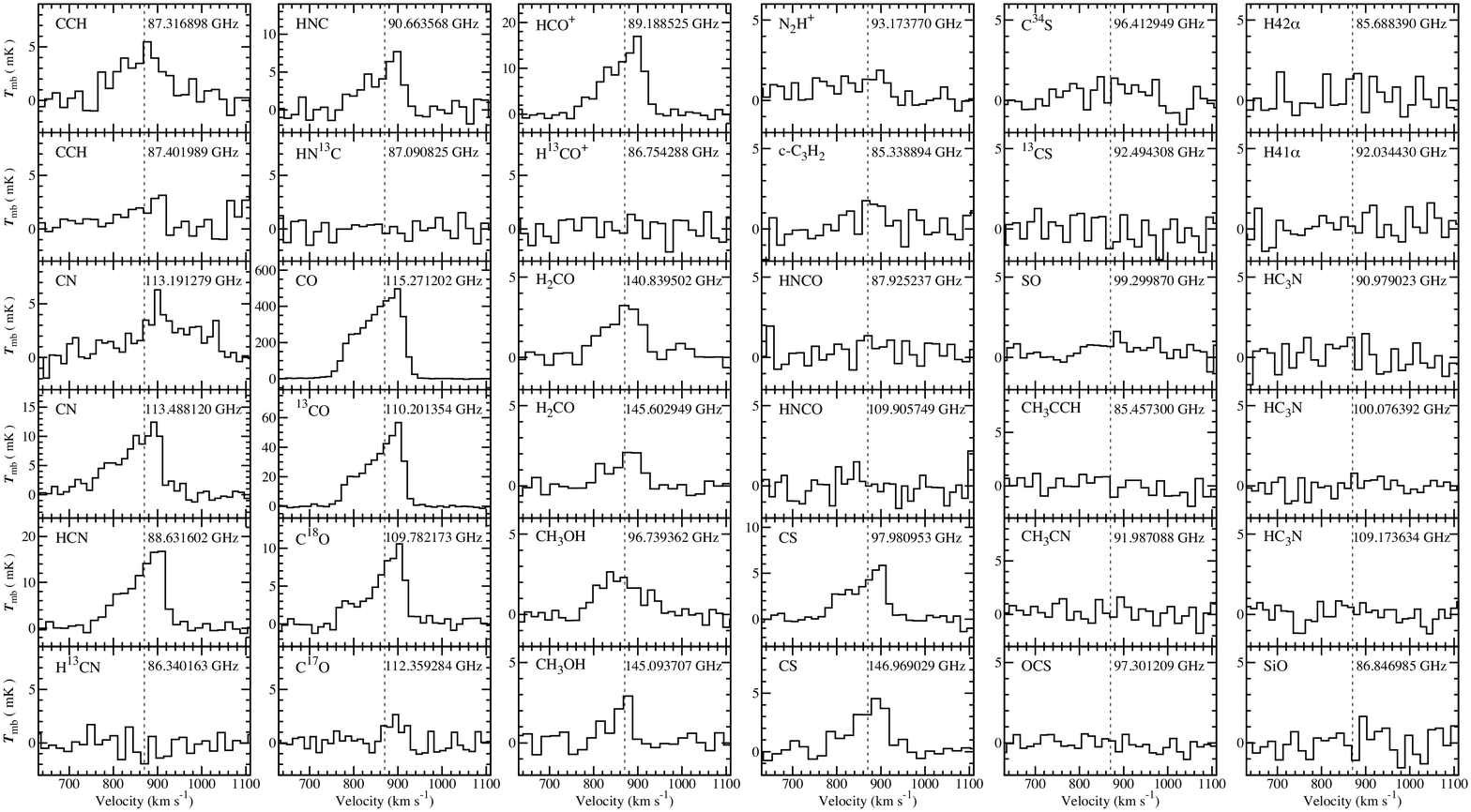}
\caption{Line profile of each molecular transition observed in BE.  Vertical dashed lines indicate $V_{\rm LSR}$ of 870~km~s$^{-1}$, which is the typical systemic velocity of BE. }
\label{fig06}
\end{figure}

\begin{figure}
\epsscale{0.99}
\plotone{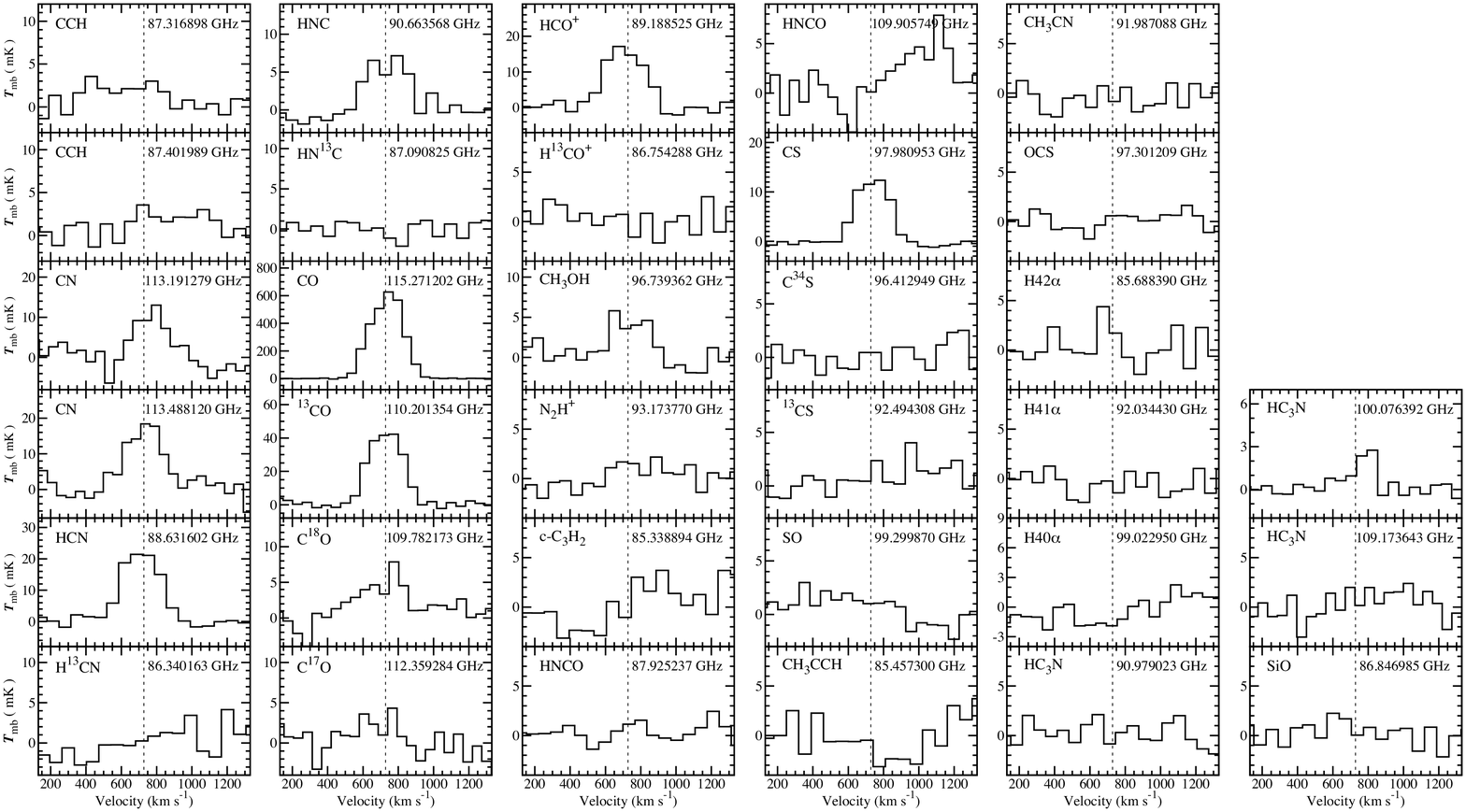}
\caption{Line profile of each molecular transition observed in NR.  Vertical dashed lines indicate $V_{\rm LSR}$ of 728~km~s$^{-1}$, which is the typical systemic velocity of NR. }
\label{fig07}
\end{figure}

\begin{figure}
\epsscale{0.99}
\plotone{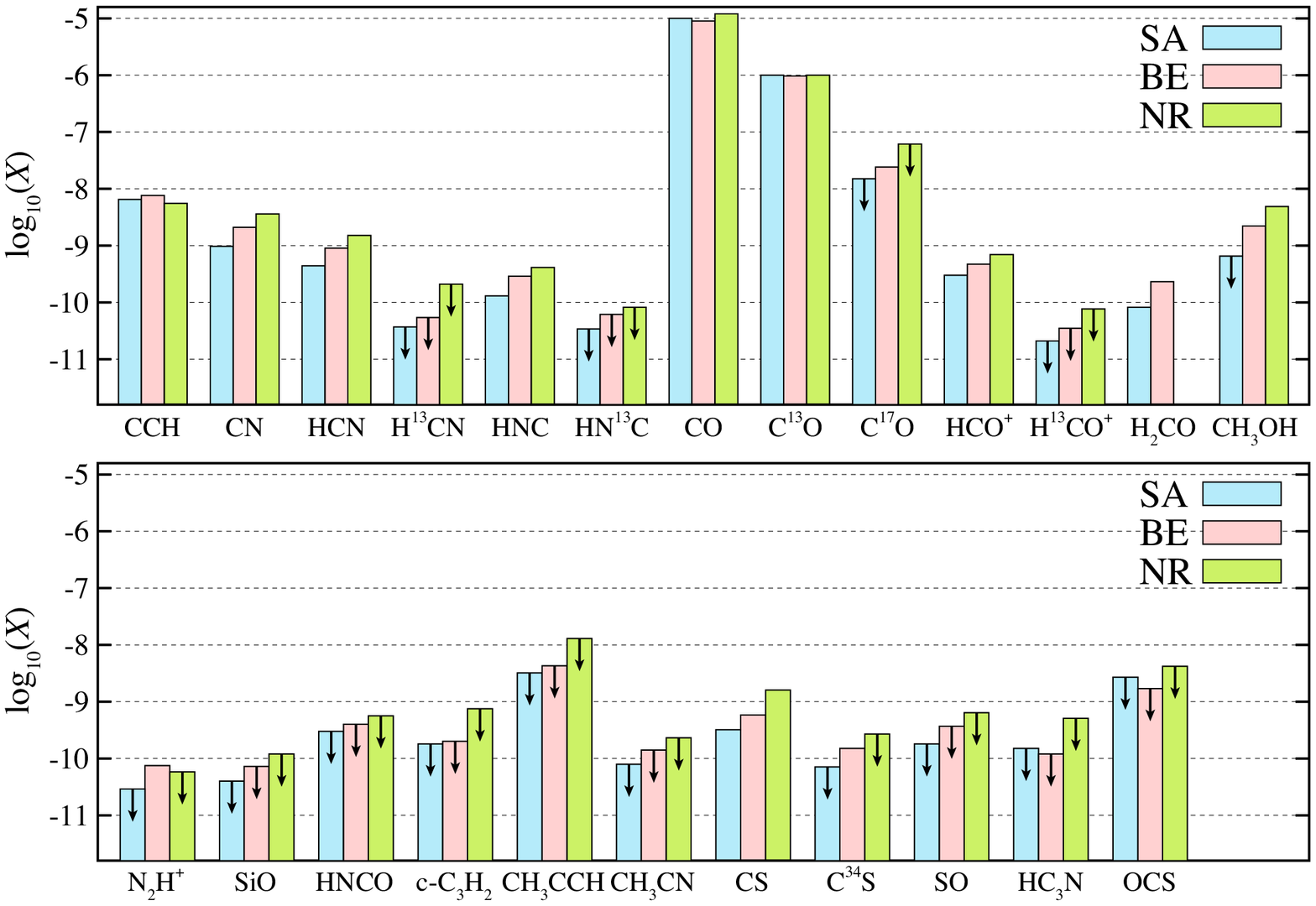}
\caption{ Fractional abundances of various molecular species relative to H$_2$ in NGC~3627~SA (blue), NGC~3627~BE (pink), and NGC~3627~NR (green).  The fractional abundances are calculated by assuming the rotation temperature of 10~K for all the positions.  Down arrows indicate upper limits.} 
\label{fig08}
\end{figure}

\begin{figure}
\epsscale{0.99}
\plotone{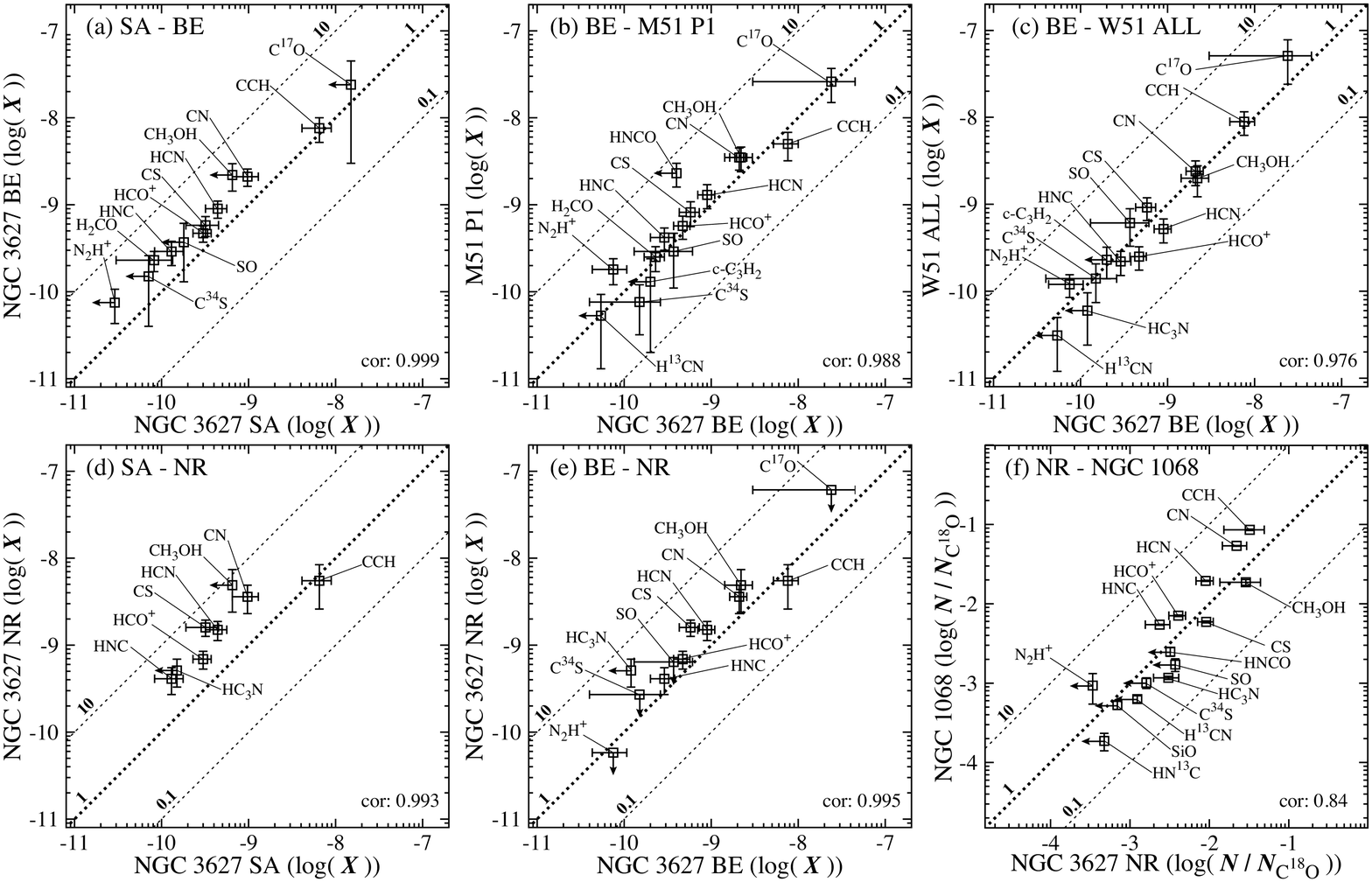}
\caption{Correlation plots of fractional abundances relative to H$_2$ ($X$) (a) between NGC~3627~SA and NGC~3627~BE, (b) between NGC~3627~BE and M51~P1, (c) between NGC~3627~BE and W51, (d) between NGC~3627~SA and NGC~3627~NR, and (e) between NGC~3627~BE and NGC~3627~NR.  (f) Correlation plot of the fractional abundances relative to C$^{18}$O between NGC~3627~NR and NGC~1068.  The fractional abundances are calculated by assuming the rotation temperature of 10~K for all the positions of NGC~3627.  The rotation temperature is assumed to be 5~K for M51~P1 \citep{Watanabe2014}.  Dashed lines indicate the fractional abundance ratios of 0.1, 1, and 10.  Arrows indicate the upper limits to the fractional abundances.  
The Pearson correlation coefficient (cor.) between $\log(X_{i,{\rm A}})$ and $\log(X_{i,{\rm B}})$ is shown in the bottom-right corner.  The upper limit values are excluded from the calculation of the correlation coefficients.}
\label{fig09}
\end{figure}

\begin{figure}
\epsscale{0.99}
\plotone{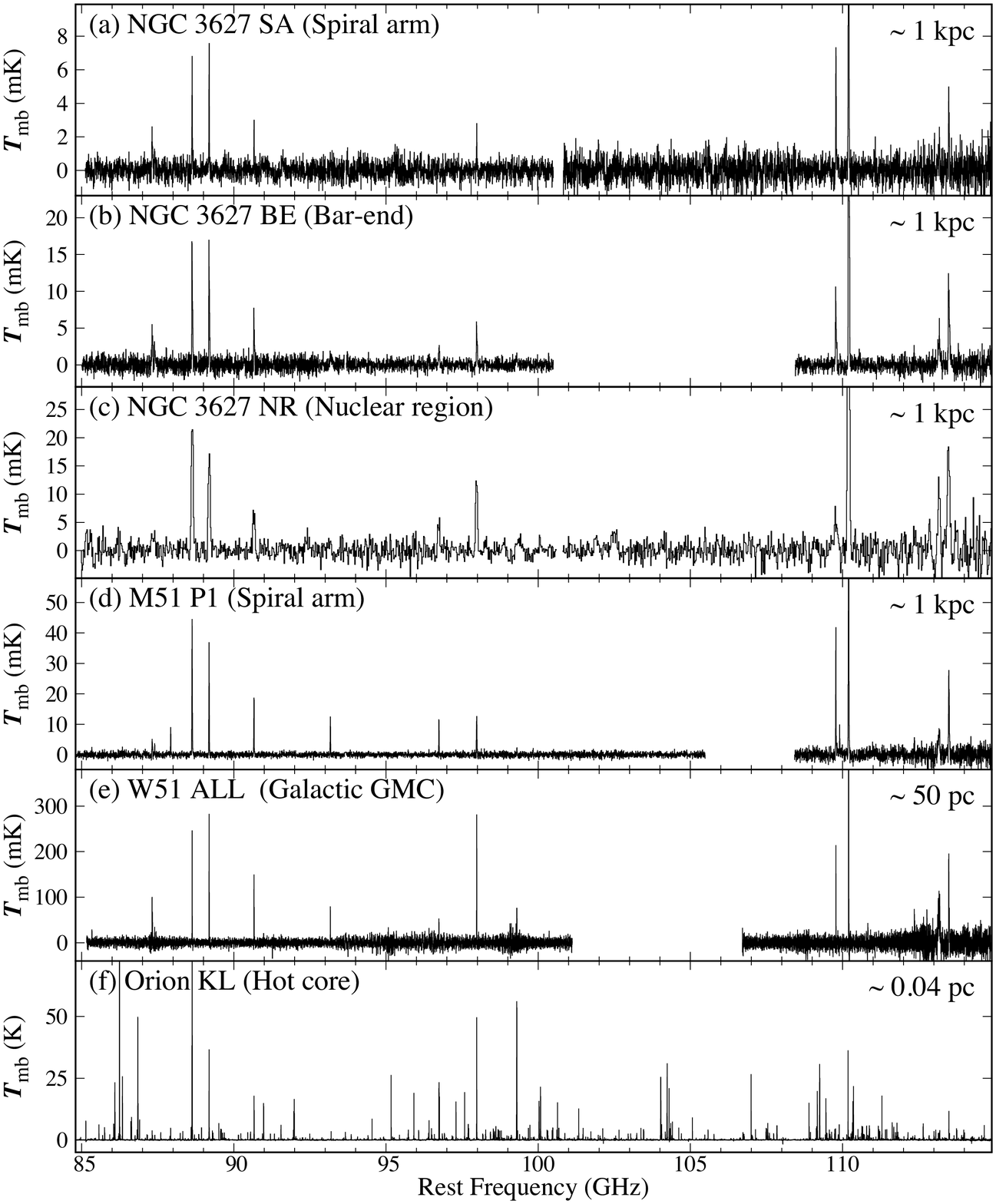}
\caption{Spectra of (a) NGC~3627~SA, (b) NGC~3627~BE, (c) NGC~3627~NR, (d) M~51~P1 \citep{Watanabe2014}, (e) W51 \citep{Watanabe2017}, and (f) Orion~KL \citep{Watanabe2015} in the 3~mm band.  The beam sizes of the observations in the linear scale are shown in the right side of panels.}
\label{fig10}
\end{figure}

\clearpage
\begin{table}
\caption{Observation Positions \label{tab01}}
\begin{tabular}{llll}
\tableline\tableline
Position  & R.A. (J2000) & Dec. (J2000) & $V_{\rm sys}$ (km~s$^{-1}$)$^{\rm a}$  \\
\tableline
Spiral Arm (SA)     & 11$^{\rm h}$ 20$^{\rm m}$ 12.1$^{\rm s}$ & +12$^{\circ}$ 59$'$ 47.1$''$ & 655 \\
Bar-End (BE)        & 11$^{\rm h}$ 20$^{\rm m}$ 16.6$^{\rm s}$ & +12$^{\circ}$ 58$'$ 44.6$''$ & 870 \\
Nuclear Region (NR) & 11$^{\rm h}$ 20$^{\rm m}$ 15.3$^{\rm s}$ & +12$^{\circ}$ 59$'$ 29.6$''$ & 715 \\
\tableline
\end{tabular}
\tablenotetext{a}{The systemic velocity assumed in the observations \citep{Watanabe2011}.}
\end{table}

\begin{table}
\caption{Surface Density of Molecular Gas, and Star Formation Rate, and Star Formation Efficiency \label{tab02}}
\begin{tabular}{llll}
\tableline\tableline
Position   & $\Sigma_{\rm SFR}$ (M$_{\odot}\,{\rm yr^{-1}} \, {\rm kpc^{-2}}$) & $\Sigma_{\rm gas}$ (M$_{\odot}\,{\rm kpc^{-2}}$) & SFE (yr$^{-1}$)  \\
\tableline
Spiral Arm (SA) $^{\rm a}$    & $0.035 \pm 0.007$ & $(2.7 \pm 0.7) \times 10^7$ & $(1.3 \pm 0.4) \times 10^{-9}$ \\
Bar-End (BE) $^{\rm a}$       & $0.31  \pm 0.06$  & $(5.5 \pm 1.3) \times 10^7$ & $(5.7 \pm 1.7) \times 10^{-9}$ \\
Nuclear Region (NR)$^{\rm b}$ & $0.20  \pm 0.04$  & $(1.1 \pm 0.6) \times 10^8$ & $(1.8 \pm 1.0) \times 10^{-9}$ \\
\tableline
\end{tabular}
\tablenotetext{a}{The values are evaluated by using the beam size of $22''$ (IRAM 30~m). A correction for the source size of $10''$ is not applied.}
\tablenotetext{b}{The values are evaluated by using the beam size of $16''$ (Nobeyama 45~m). A correction for the source size of $10''$ is not applied.}
\end{table}

\begin{table}
\small
\caption{Observation Settings \label{tab03}}
\begin{tabular}{llllll}
\tableline\tableline
\multicolumn{6}{c}{Spiral Arm (SA)} \\ 
\tableline
  & Telescope & Receiver & LSB (GHz) & USB (GHz) & $T_{\rm sys}$ (K) $^{\rm a}$ \\
\tableline
Set 1 & IRAM~30~m & E090   &  85.1 --  92.9 & 100.8 -- 108.6 & 60 -- 140 \\
Set 2 & IRAM~30~m & E090   &  85.1 --  92.9 & N/A            & 70 -- 110 \\
      &           & E150   & 140.0 -- 147.8 & N/A            & 80 -- 140 \\
Set 3 & IRAM~30~m & E090   &  92.7 -- 100.5 & 108.4 -- 116.2 & 60 -- 190 \\
\tableline
\multicolumn{6}{c}{Bar-End (BE)} \\ 
\tableline
  & Telescope & Receiver & LSB (GHz) & USB (GHz) & $T_{\rm sys}$ (K) $^{\rm a}$ \\
\tableline
Set 1 & IRAM~30~m & E090   &  85.0 --  92.8 & N/A            & 80 -- 110 \\
      &           & E150   & 140.0 -- 147.8 & N/A            & 90 -- 150 \\
Set 2 & IRAM~30~m & E090   &  92.7 -- 100.5 & 108.4 -- 116.2 & 70 -- 290 \\
\tableline
\multicolumn{6}{c}{Nuclear Region (NR)} \\ 
\tableline
  & Telescope & Receiver & LSB (GHz) & USB (GHz) & $T_{\rm sys}$ (K) $^{\rm a}$ \\
\tableline
Set 1 & Nobeyama~45~m  & TZ1H/V &  85.0 --  88.1 &  97.0 -- 100.3 & 140 -- 250 \\
Set 2 & Nobeyama~45~m  & TZ1H/V &  88.0 --  91.1 &  99.8 -- 103.1 & 110 -- 180 \\
Set 3 & Nobeyama~45~m  & TZ1H/V &  91.0 --  94.1 & 103.0 -- 106.1 & 110 -- 260 \\
Set 4 & Nobeyama~45~m  & TZ1H/V &  94.0 --  97.1 & 106.0 -- 109.1 & 140 -- 250 \\
Set 5 & Nobeyama~45~m  & TZ1H/V &  97.0 -- 100.3 & 109.0 -- 112.1 & 130 -- 160 \\
Set 6 & Nobeyama~45~m  & TZ1H/V &  99.8 -- 103.1 & 112.1 -- 115.5 & 150 -- 210 \\
\tableline
\end{tabular}
\tablenotetext{a}{The system noise temperature during the observation run.}
\end{table}

\begin{table}
\small
\caption{Summary of Spectra \label{tab04}}
\begin{tabular}{lllll}
\tableline\tableline
Position  & Band & Frequency & Resolution & Sensitivity  \\
          &      & (GHz)     & (MHz)      & (mK)         \\
\tableline
NGC~3627~SA  & 3~mm & 85.1 -- 100.5, 100.8 -- 116.2 &  5 & 0.4 -- 1.3  \\
             & 2~mm & 140.0 -- 147.8                & 10 & 0.4         \\
NGC~3627~BE  & 3~mm & 85.1 -- 100.5, 108.4 -- 116.2 &  5 & 0.5 -- 1.6  \\
             & 2~mm & 140.0 -- 147.8                & 10 & 0.5 -- 0.6  \\
NGC~3627~NR  & 3~mm & 85.0 -- 115.5                 & 20 & 0.8 -- 3.2  \\
\tableline
\end{tabular}
\end{table}

\begin{table}
\tiny 
\caption{Line Parameters in NGC~3627~SA \label{tab05}}
\begin{tabular}{lllllllll}
\tableline\tableline
               & Transition & Freq. & $E_{\rm u}$ & S$\mu^2$ & $T_{\rm mb}$ Peak \tablenotemark{ab} & $\int T_{\rm mb} dv$ \tablenotemark{ac} & $V_{\rm LSR}$ \tablenotemark{d} & FWHM \tablenotemark{d} \\
               &            & (GHz) & (K)        &           & (K)               & (K km s$^{-1}$)      & (km s$^{-1}$) & (km s$^{-1}$) \\
\tableline
c-C$_3$H$_2$               & $2_{1\,2}-1_{0\,1}$         &  85.338894 &  4.1 \tablenotemark{e} & 16.1 \tablenotemark{e}  & $< 0.001$ & $< 0.04$   & \nodata           & \nodata    \\
CH3CCH                     & $5_0-4_0$                   &  85.457300 & 12.3                   & 6.15                    & $< 0.001$ & $< 0.04$   & \nodata           & \nodata    \\
H42$\alpha$                &                             &  85.688390 &                        &                         & $< 0.001$ & $< 0.04$   & \nodata           & \nodata    \\
H$^{13}$CN                 & $1-0$                       &  86.340163 &  4.1                   & 8.91 \tablenotemark{g}  & $< 0.001$ & $< 0.03$   & \nodata           & \nodata    \\
H$^{13}$CO$^+$             & $1-0$                       &  86.754288 &  4.2                   & 15.2                    & $< 0.001$ & $< 0.03$   & \nodata           & \nodata    \\
SiO                        & $2-1$                       &  86.846985 &  6.3                   & 19.2                    & $< 0.001$ & $< 0.03$   & \nodata           & \nodata    \\
HN$^{13}$C                 & $1-0$                       &  87.090825 &  4.2                   & 9.30 \tablenotemark{g}  & $< 0.001$ & $< 0.03$   & \nodata           & \nodata    \\
CCH \tablenotemark{h}      & $N=1-0,J=3/2-1/2,F=2-1$     &  87.316898 &  4.2                   & 0.988                   & 0.003 (1) &  0.23 (7)  & \tablenotemark{i} & \nodata    \\
CCH \tablenotemark{h}      & $N=1-0,J=3/2-1/2,F=1-0$     &  87.328585 &  4.2                   & 0.492                   &           &            & \nodata           & \nodata    \\
CCH \tablenotemark{h}      & $N=1-0,J=1/2-1/2,F=1-1$     &  87.401989 &  4.2                   & 0.492                   & 0.001 (1) &  0.10 (6)  & \tablenotemark{i} & \nodata    \\
CCH \tablenotemark{h}      & $N=1-0,J=1/2-1/2,F=0-1$     &  87.407165 &  4.2                   & 0.198                   &           &            & \nodata           & \nodata    \\
HNCO                       & $4_{0\,4}-3_{0\,3}$         &  87.925237 & 10.5                   & 9.99 \tablenotemark{g}  & $< 0.001$ & $< 0.04$   & \nodata           & \nodata    \\
HCN                        & $1-0$                       &  88.631602 &  4.3                   & 8.91 \tablenotemark{g}  & 0.007 (2) &  0.39 (7)  & 663 (2)           & 58 (4)     \\
HCO$^+$                    & $1-0$                       &  89.188525 &  4.3                   & 15.2                    & 0.008 (1) &  0.46 (6)  & 658 (1)           & 59 (3)     \\
HNC                        & $1-0$                       &  90.663568 &  4.4                   & 9.30 \tablenotemark{g}  & 0.003 (1) &  0.13 (4)  & 657 (2)           & 41 (5)     \\
HC$_3$N                    & $10-9$                      &  90.979023 & 24.0                   & 139.3 \tablenotemark{g} & $< 0.001$ & $< 0.04$   & \nodata           & \nodata    \\
CH$_3$CN                   & $5_0-4_0$                   &  91.987088 & 13.2                   & 153.8                   & $< 0.001$ & $< 0.03$   & \nodata           & \nodata    \\
H41$\alpha$                &                             &  92.034430 &                        &                         & $< 0.001$ & $< 0.03$   & \nodata           & \nodata    \\
$^{13}$CS                  & $2-1$                       &  92.494308 &  6.7                   & 15.3                    & $< 0.001$ & $< 0.04$   & \nodata           & \nodata    \\
N$_2$H$^+$                 & $1-0$                       &  93.173770 &  4.5                   & 104.1                   & $< 0.002$ & $< 0.04$   & \nodata           & \nodata    \\
C$^{34}$S                  & $2-1$                       &  96.412949 &  6.9                   & 7.67                    & $< 0.001$ & $< 0.03$   & \nodata           & \nodata    \\
CH$_3$OH \tablenotemark{h} & $2_{-1}-1_{-1},{E}$         &  96.739362 &  4.6 \tablenotemark{j} & 1.21                    & $< 0.001$ & $< 0.04$   & \nodata           & \nodata    \\
CH$_3$OH \tablenotemark{h} & $2_0-1_0,{\rm A}^+$         &  96.741375 &  7.0                   & 1.62                    &           &            & \nodata           & \nodata    \\
CH$_3$OH \tablenotemark{h} & $2_{0}-1_{0},{E}$           &  96.744550 & 12.2 \tablenotemark{j} & 1.62                    &           &            & \nodata           & \nodata    \\
OCS                        & $8-7$                       &  97.301209 & 21.0                   & 4.09                    & $< 0.001$ & $< 0.08$   & \nodata           & \nodata    \\
CS                         & $2-1$                       &  97.980953 &  7.1                   & 7.67                    & 0.003 (1) &  0.14 (5)  & 661 (3)           & 47 (7)     \\
H40$\alpha$                &                             &  99.022950 &                        &                         & $< 0.001$ & $< 0.03$   & \nodata           & \nodata    \\
SO                         & $J_N=3_2-2_1$               &  99.299870 &  9.2                   & 6.91                    & $< 0.001$ & $< 0.03$   & \nodata           & \nodata    \\
HC$_3$N                    & $11-10$                     & 100.076392 & 28.8                   & 153.2 \tablenotemark{g} & $< 0.001$ & $< 0.03$   & \nodata           & \nodata    \\
HC$_3$N                    & $12-11$                     & 109.173634 & 34.1                   & 167.1 \tablenotemark{g} & $< 0.002$ & $< 0.04$   & \nodata           & \nodata    \\
OCS                        & $9-8$                       & 109.463063 & 26.3                   & 4.60                    & $< 0.002$ & $< 0.04$   & \nodata           & \nodata    \\
C$^{18}$O                  & $1-0$                       & 109.782173 &  5.3                   & 0.0122                  & 0.007 (2) &  0.42 (7)  & 657 (1)           & 49 (3)     \\
HNCO                       & $5_{0\,5}-4_{0\,4}$         & 109.905749 & 15.8                   & 12.5 \tablenotemark{g}  & $< 0.005$ & $< 0.04$   & \nodata           & \nodata    \\
$^{13}$CO                  & $1-0$                       & 110.201354 &  5.3                   & 0.0122                  & 0.047 (2) &  2.50 (7)  & 661.3 (3)         & 50.2 (7) \\
C$^{17}$O                  & $1-0$                       & 112.359284 &  5.4                   & 0.0122                  & $< 0.002$ & $< 0.04$   & \nodata           & \nodata    \\
CN \tablenotemark{h}       & $N=1-0,J=1/2-1/2,F=1/2-3/2$ & 113.144157 &  5.4                   & 1.25                    & 0.003 (2) &  0.15 (8)  & \tablenotemark{i} & \nodata    \\
CN \tablenotemark{h}       & $N=1-0,J=1/2-1/2,F=3/2-1/2$ & 113.170492 &  5.4                   & 1.22                    &           &            & \nodata           & \nodata    \\
CN \tablenotemark{h}       & $N=1-0,J=1/2-1/2,F=3/2-3/2$ & 113.191279 &  5.4                   & 1.58                    &           &            & \nodata           & \nodata    \\
CN \tablenotemark{h}       & $N=1-0,J=3/2-1/2,F=5/2-3/2$ & 113.490970 &  5.4                   & 1.58                    & 0.005 (2) &  0.30 (8)  & \tablenotemark{i} & \nodata    \\
CN \tablenotemark{h}       & $N=1-0,J=3/2-1/2,F=3/2-1/2$ & 113.488120 &  5.4                   & 4.21                    &           &            & \nodata           & \nodata    \\
CN \tablenotemark{h}       & $N=1-0,J=3/2-1/2,F=1/2-1/2$ & 113.499644 &  5.4                   & 1.25                    &           &            & \nodata           & \nodata    \\
CO                         & $1-0$                       & 115.271202 &  5.5                   & 0.0121                  & 0.498 (4) & 29.7 (2)   & 662.9 (3)         & 52.7 (7) \\
H$_2$CO                    & $2_{1\,2}-1_{1\,1}$         & 140.839502 &  6.7 \tablenotemark{k} & 8.16                    & 0.0014 (9)&  0.05 (3)  & \tablenotemark{l} & \nodata    \\
CH$_3$OH \tablenotemark{h} & $3_0-2_0,{\rm E}$           & 145.093707 & 19.2 \tablenotemark{j} & 2.42                    & 0.001 (1) &  0.06 (5)  & \tablenotemark{l} & \nodata    \\
CH$_3$OH \tablenotemark{h} & $3_{-1}-2_{-1},{\rm E}$     & 145.097370 & 11.6 \tablenotemark{j} & 2.16                    &           &            & \nodata           & \nodata    \\
CH$_3$OH \tablenotemark{h} & $3_0-2_0,{\rm A}^+$         & 145.103152 & 13.9                   & 2.43                    &           &            & \nodata           & \nodata    \\
H$_2$CO                    & $2_{0\,2}-1_{0\,1}$         & 145.602949 & 10.5                   & 10.9                    & $< 0.001$ & $< 0.04$   & \nodata           & \nodata    \\
CS                         & $3-2$                       & 146.969029 & 14.1                   & 11.5                    & 0.001 (1) &  0.06 (6)  & 657 (7)           & 40 (20)    \\
\tableline
\end{tabular}
\tablenotetext{a}{The numbers in parentheses represent $3\sigma$ errors.}
\tablenotetext{b}{Upper limits to the peak temperature are $3\sigma$.}
\tablenotetext{c}{The upper limit to the integrated intensity is calculated as: $\int T_{\rm mb} dv < 3 \sigma \times \sqrt{\Delta V \times \Delta v_{\rm res}}$, where $\Delta V$ is the assumed line width (50~km~s$^{-1}$) and $\Delta v_{\rm res}$ is the velocity resolution per channel.}
\tablenotetext{d}{The numbers in parentheses represent $1\sigma$ errors in units of the last significant digits.}
\tablenotetext{e}{The upper state energy is calculated from the lowest ortho state ($1_{0\,1}$).}
\tablenotetext{f}{The spin weight of 3 is not included.}
\tablenotetext{g}{The nuclear spin multiplicity of 3 for the N nucleus is not included.}
\tablenotetext{h}{The line is blended with other lines.}
\tablenotetext{i}{Gaussian fitting is not successful due to blending with other lines.}
\tablenotetext{j}{The upper state energy is calculated from the lowest E state ($1_{-1}$, E).}
\tablenotetext{k}{The upper state energy is calculated from the lowest ortho state ($1_{1\,1}$).}
\tablenotetext{l}{Gaussian fitting is not successful due to the poor S/N ratio.}
\end{table}

\begin{table}
\tiny
\caption{Line Parameters in NGC~3627~BE \label{tab06}}
\begin{tabular}{lllllllll}
\tableline\tableline
               & Transition & Freq. & $E_{\rm u}$ & S$\mu^2$ & $T_{\rm mb}$ Peak \tablenotemark{ab} & $\int T_{\rm mb} dv$ \tablenotemark{ac} & $V_{\rm LSR}$ \tablenotemark{d} & FWHM \tablenotemark{d} \\
               &            & (GHz) & (K)        &           & (K)               & (K km s$^{-1}$)      & (km s$^{-1}$) & (km s$^{-1}$) \\
\tableline
c-C$_3$H$_2$               & $2_{1\,2}-1_{0\,1}$         &  85.338894 &  4.1 \tablenotemark{e} & 16.1 \tablenotemark{e}   & $< 0.002$ & $< 0.09$  & \nodata           & \nodata  \\
CH3CCH                     & $5_0-4_0$                   &  85.457300 & 12.3                   & 6.15                     & $< 0.003$ & $< 0.1$   & \nodata           & \nodata  \\
H42$\alpha$                &                             &  85.688390 &                        &                          & $< 0.002$ & $< 0.09$  & \nodata           & \nodata  \\
H$^{13}$CN                 & $1-0$                       &  86.340163 &  4.1                   & 8.91 \tablenotemark{g}   & $< 0.002$ & $< 0.09$  & \nodata           & \nodata  \\
H$^{13}$CO$^+$             & $1-0$                       &  86.754288 &  4.2                   & 15.2                     & $< 0.002$ & $< 0.1$   & \nodata           & \nodata  \\
SiO                        & $2-1$                       &  86.846985 &  6.3                   & 19.2                     & $< 0.003$ & $< 0.1$   & \nodata           & \nodata  \\
HN$^{13}$C                 & $1-0$                       &  87.090825 &  4.2                   & 9.30 \tablenotemark{g}   & $< 0.003$ & $< 0.1$   & \nodata           & \nodata  \\
CCH \tablenotemark{h}      & $N=1-0,J=3/2-1/2,F=2-1$     &  87.316898 &  4.2                   & 0.988                    & 0.006 (2) &  0.6 (1)  & \tablenotemark{i} & \nodata  \\
CCH \tablenotemark{h}      & $N=1-0,J=3/2-1/2,F=1-0$     &  87.328585 &  4.2                   & 0.492                    &           &           & \nodata           & \nodata  \\
CCH \tablenotemark{h}      & $N=1-0,J=1/2-1/2,F=1-1$     &  87.401989 &  4.2                   & 0.492                    & 0.003 (2) &  0.3 (1)  & \tablenotemark{i} & \nodata  \\
CCH \tablenotemark{h}      & $N=1-0,J=1/2-1/2,F=0-1$     &  87.407165 &  4.2                   & 0.198                    &           &           & \nodata           & \nodata  \\
HNCO                       & $4_{0\,4}-3_{0\,3}$         &  87.925237 & 10.5                   & 9.99 \tablenotemark{g}   & $< 0.003$ & $< 0.1$   & \nodata           & \nodata  \\
HCN                        & $1-0$                       &  88.631602 &  4.3                   & 8.91 \tablenotemark{g}   & 0.017 (3) &  1.6 (2)  & 876 (4)           & 102 (8)  \\
HCO$^+$                    & $1-0$                       &  89.188525 &  4.3                   & 15.2                     & 0.017 (3) &  1.5 (2)  & 872 (3)           & 101 (8)  \\
HNC                        & $1-0$                       &  90.663568 &  4.4                   & 9.30 \tablenotemark{g}   & 0.008 (3) &  0.6 (1)  & 872 (6)           &  90 (10) \\
HC$_3$N                    & $10-9$                      &  90.979023 & 24.0                   & 139.3 \tablenotemark{g}  & $< 0.002$ &  $< 0.09$ & \nodata           & \nodata  \\
CH$_3$CN                   & $5_0-4_0$                   &  91.987088 & 13.2                   & 153.8                    & $< 0.003$ &  $< 0.1$  & \nodata           & \nodata  \\
H41$\alpha$                &                             &  92.034430 &                        &                          & $< 0.003$ &  $< 0.1$  & \nodata           & \nodata  \\
$^{13}$CS                  & $2-1$                       &  92.494308 &  6.7                   & 15.3                     & $< 0.003$ &  $< 0.1$  & \nodata           & \nodata  \\
N$_2$H$^+$                 & $1-0$                       &  93.173770 &  4.5                   & 104.1                    & 0.002 (1) &  0.21 (8) & \tablenotemark{j} & \nodata  \\
C$^{34}$S                  & $2-1$                       &  96.412949 &  6.9                   & 7.67                     & 0.002 (2) &  0.13 (9) & \tablenotemark{j} & \nodata  \\
CH$_3$OH \tablenotemark{h} & $2_{-1}-1_{-1},{E}$         &  96.739362 &  4.6 \tablenotemark{k} & 1.21                     & 0.003 (2) &  0.28 (8) & 863 (5)           & 110 (10) \\
CH$_3$OH \tablenotemark{h} & $2_0-1_0,{\rm A}^+$         &  96.741375 &  7.0                   & 1.62                     &           &           & \nodata           & \nodata  \\
CH$_3$OH \tablenotemark{h} & $2_{0}-1_{0},{E}$           &  96.744550 & 12.2 \tablenotemark{k} & 1.62                     &           &           & \nodata           & \nodata  \\
OCS                        & $8-7$                       &  97.301209 & 21.0                   & 4.09                     & $< 0.001$ &  $< 0.05$ & \nodata           & \nodata  \\
CS                         & $2-1$                       &  97.980953 &  7.1                   & 7.67                     & 0.006 (2) &  0.52 (9) & 868 (5)           & 110 (10) \\
H40$\alpha$                &                             &  99.022950 &                        &                          & $< 0.002$ &  $< 0.06$ & \nodata           & \nodata  \\
SO                         & $J_N=3_2-2_1$               &  99.299870 &  9.2                   & 6.91                     & 0.002 (2) &  0.13 (8) &                   & \nodata  \\
HC$_3$N                    & $11-10$                     & 100.076392 & 28.8                   & 153.2 \tablenotemark{g}  & $< 0.001$ &  $< 0.05$ & \nodata           & \nodata  \\
HC$_3$N                    & $12-11$                     & 109.173634 & 34.1                   & 167.1 \tablenotemark{g}  & $< 0.002$ &  $< 0.06$ & \nodata           & \nodata  \\
OCS                        & $9-8$                       & 109.463063 & 26.3                   & 4.60                     & $< 0.002$ &  $< 0.06$ & \nodata           & \nodata  \\
C$^{18}$O                  & $1-0$                       & 109.782173 &  5.3                   & 0.0122                   & 0.011 (2) &  0.9 (1)  & 883 (3)           &  82 (7)  \\
HNCO                       & $5_{0\,5}-4_{0\,4}$         & 109.905749 & 15.8                   & 12.5 \tablenotemark{g}   & $< 0.008$ &  $< 0.07$ & \nodata           & \nodata  \\
$^{13}$CO                  & $1-0$                       & 110.201354 &  5.3                   & 0.0122                   & 0.057 (2) &  4.9 (1)  & 870 (3)           & 103 (7)  \\
C$^{17}$O                  & $1-0$                       & 112.359284 &  5.4                   & 0.0122                   & 0.003 (2) &  0.1 (1)  & 892 (5)           &  40 (10) \\
CN \tablenotemark{h}       & $N=1-0,J=1/2-1/2,F=1/2-3/2$ & 113.144157 &  5.4                   & 1.25                     & 0.006 (3) &  0.7 (2)  & \tablenotemark{i} & \nodata  \\
CN \tablenotemark{h}       & $N=1-0,J=1/2-1/2,F=3/2-1/2$ & 113.170492 &  5.4                   & 1.22                     &           &           & \nodata           & \nodata  \\
CN \tablenotemark{h}       & $N=1-0,J=1/2-1/2,F=3/2-3/2$ & 113.191279 &  5.4                   & 1.58                     &           &           & \nodata           & \nodata  \\
CN \tablenotemark{h}       & $N=1-0,J=3/2-1/2,F=5/2-3/2$ & 113.490970 &  5.4                   & 1.58                     & 0.012 (3) &  1.3 (1)  & \tablenotemark{i} & \nodata  \\
CN \tablenotemark{h}       & $N=1-0,J=3/2-1/2,F=3/2-1/2$ & 113.488120 &  5.4                   & 4.21                     &           &           & \nodata           & \nodata  \\
CN \tablenotemark{h}       & $N=1-0,J=3/2-1/2,F=1/2-1/2$ & 113.499644 &  5.4                   & 1.25                     &           &           & \nodata           & \nodata  \\
CO                         & $1-0$                       & 115.271202 &  5.5                   & 0.0121                   & 0.497 (4) & 52.9 (2)  & 862 (3)           & 115 (6)  \\
H$_2$CO                    & $2_{1\,2}-1_{1\,1}$         & 140.839502 &  6.7 \tablenotemark{l} & 8.16                     & 0.003 (1) &  0.31 (8) & 868 (5)           & 100 (10) \\
CH$_3$OH \tablenotemark{h} & $3_0-2_0,{\rm E}$           & 145.093707 & 19.2 \tablenotemark{k} & 2.42                     & 0.003 (1) &  0.15 (8) & 877 (3)           &  34 (9)  \\
CH$_3$OH \tablenotemark{h} & $3_{-1}-2_{-1},{\rm E}$     & 145.097370 & 11.6 \tablenotemark{k} & 2.16                     &           &           & \nodata           & \nodata  \\
CH$_3$OH \tablenotemark{h} & $3_0-2_0,{\rm A}^+$         & 145.103152 & 13.9                   & 2.43                     &           &           & \nodata           & \nodata  \\
H$_2$CO                    & $2_{0\,2}-1_{0\,1}$         & 145.602949 & 10.5                   & 10.9                     & 0.002 (1) &  0.17 (8) & 874 (7)           &  80 (20) \\
CS                         & $3-2$                       & 146.969029 & 14.1                   & 11.5                     & 0.005 (1) &  0.42 (6) & 878 (4)           &  90 (10) \\
\tableline
\end{tabular}
\tablenotetext{a}{The numbers in parentheses represent $3\sigma$ errors.}
\tablenotetext{b}{Upper limits to the peak temperature are $3\sigma$.}
\tablenotetext{c}{The upper limit to the integrated intensity is calculated as: $\int T_{\rm mb} dv < 3 \sigma \times \sqrt{\Delta V \times \Delta v_{\rm res}}$, where $\Delta V$ is the assumed line width (100~km~s$^{-1}$) and $\Delta v_{\rm res}$ is the velocity resolution per channel.}
\tablenotetext{d}{The numbers in parentheses represent $1\sigma$ errors in units of the last significant digits.}
\tablenotetext{e}{The upper state energy is calculated from the lowest ortho state ($1_{0\,1}$).}
\tablenotetext{f}{The spin weight of 3 is not included.}
\tablenotetext{g}{The nuclear spin multiplicity of 3 for the N nucleus is not included.}
\tablenotetext{h}{The line is blended with other lines.}
\tablenotetext{i}{Gaussian fitting is not successful due to blending with other lines.}
\tablenotetext{j}{Gaussian fitting is not successful due to the poor S/N ratio.}
\tablenotetext{k}{The upper state energy is calculated from the lowest E state ($1_{-1}$, E).}
\tablenotetext{l}{The upper state energy is calculated from the lowest ortho state ($1_{1\,1}$).}
\end{table}

\begin{table}
\tiny
\caption{Line Parameters in NGC~3627~NR \label{tab07}}
\begin{tabular}{lllllllll}
\tableline\tableline
               & Transition & Freq. & $E_{\rm u}$ & S$\mu^2$ & $T_{\rm mb}$ Peak \tablenotemark{ab} & $\int T_{\rm mb} dv$ \tablenotemark{ac} & $V_{\rm LSR}$ \tablenotemark{d} & FWHM \tablenotemark{d} \\
                           &                             & (GHz) & (K)        &           & (K)               & (K km s$^{-1}$)      & (km s$^{-1}$) & (km s$^{-1}$) \\
\tableline
c-C$_3$H$_2$               & $2_{1\,2}-1_{0\,1}$         &  85.338894 &  4.1 \tablenotemark{e} & 16.1 \tablenotemark{f}  & $<0.006$  &  $< 0.7$  & \nodata           & \nodata  \\
CH3CCH                     & $5_0-4_0$                   &  85.457300 & 12.3                   & 6.15                    & $<0.006$  &  $< 0.7$  & \nodata           & \nodata  \\
H42$\alpha$                &                             &  85.688390 &                        &                         & $<0.004$  &  $< 0.5$  & \nodata           & \nodata  \\
H$^{13}$CN                 & $1-0$                       &  86.340163 &  4.1                   & 8.91 \tablenotemark{g}  & $<0.006$  &  $< 0.6$  & \nodata           & \nodata  \\
H$^{13}$CO$^+$             & $1-0$                       &  86.754288 &  4.2                   & 15.2                    & $<0.004$  &  $< 0.5$  & \nodata           & \nodata  \\
SiO                        & $2-1$                       &  86.846985 &  6.3                   & 19.2                    & $<0.003$  &  $< 0.4$  & \nodata           & \nodata  \\
HN$^{13}$C                 & $1-0$                       &  87.090825 &  4.2                   & 9.30 \tablenotemark{g}  & $<0.003$  &  $< 0.3$  & \nodata           & \nodata  \\
CCH \tablenotemark{h}      & $N=1-0,J=3/2-1/2,F=2-1$     &  87.316898 &  4.2                   & 0.988                   & 0.004 (3) &   1.2 (6) & \tablenotemark{i} & \nodata  \\
CCH \tablenotemark{h}      & $N=1-0,J=3/2-1/2,F=1-0$     &  87.328585 &  4.2                   & 0.492                   &           &           & \nodata           & \nodata  \\
CCH \tablenotemark{h}      & $N=1-0,J=1/2-1/2,F=1-1$     &  87.401989 &  4.2                   & 0.492                   &           &           & \nodata           & \nodata  \\
CCH \tablenotemark{h}      & $N=1-0,J=1/2-1/2,F=0-1$     &  87.407165 &  4.2                   & 0.198                   &           &           & \nodata           & \nodata  \\
HNCO                       & $4_{0\,4}-3_{0\,3}$         &  87.925237 & 10.5                   & 9.99 \tablenotemark{g}  & $<0.003$  &  $< 0.3$  & \nodata           & \nodata  \\
HCN                        & $1-0$                       &  88.631602 &  4.3                   & 8.91 \tablenotemark{g}  & 0.021 (4) &   6.0 (8) & 710 (6)           & 240 (20) \\
HCO$^+$                    & $1-0$                       &  89.188525 &  4.3                   & 15.2                    & 0.017 (3) &   4.6 (6) & 703 (9)           & 250 (20) \\
HNC                        & $1-0$                       &  90.663568 &  4.4                   & 9.30 \tablenotemark{g}  & 0.007 (3) &   1.8 (5) & 740 (20)          & 260 (40) \\
HC$_3$N                    & $10-9$                      &  90.979023 & 24.0                   & 139.3 \tablenotemark{g} & $<0.004$  &  $< 0.5$  & \nodata           & \nodata  \\
CH$_3$CN                   & $5_0-4_0$                   &  91.987088 & 13.2                   & 153.8                   & $<0.003$  &  $< 0.4$  & \nodata           & \nodata  \\
H41$\alpha$                &                             &  92.034430 &                        &                         & $<0.003$  &  $< 0.4$  & \nodata           & \nodata  \\
$^{13}$CS                  & $2-1$                       &  92.494308 &  6.7                   & 15.3                    & $<0.003$  &  $< 0.4$  & \nodata           & \nodata  \\
N$_2$H$^+$                 & $1-0$                       &  93.173770 &  4.5                   & 104.1                   & $<0.003$  &  $< 0.3$  & \nodata           & \nodata  \\
C$^{34}$S                  & $2-1$                       &  96.412949 &  6.9                   & 7.67                    & $<0.004$  &  $< 0.5$  & \nodata           & \nodata  \\
CH$_3$OH \tablenotemark{h} & $2_{-1}-1_{-1},{E}$         &  96.739362 &  4.6 \tablenotemark{j} & 1.21                    & 0.006 (4) &   1.3 (6) & 740 (20)          & 240 (60) \\
CH$_3$OH \tablenotemark{h} & $2_0-1_0,{\rm A}^+$         &  96.741375 &  7.0                   & 1.62                    &           &           & \nodata           & \nodata  \\
CH$_3$OH \tablenotemark{h} & $2_{0}-1_{0},{E}$           &  96.744550 & 12.2 \tablenotemark{j} & 1.62                    &           &           & \nodata           & \nodata  \\
OCS                        & $8-7$                       &  97.301209 & 21.0                   & 4.09                    & $<0.002$  &  $< 0.3$  & \nodata           & \nodata  \\
CS                         & $2-1$                       &  97.980953 &  7.1                   & 7.67                    & 0.012 (1) &   2.9 (2) & 740 (6)           & 210 (10) \\
H40$\alpha$                &                             &  99.022950 &                        &                         & $<0.004$  &  $< 0.4$  & \nodata           & \nodata  \\
SO                         & $J_N=3_2-2_1$               &  99.299870 &  9.2                   & 6.91                    & $<0.004$  &  $< 0.5$  & \nodata           & \nodata  \\
HC$_3$N                    & $11-10$                     & 100.076392 & 28.8                   & 153.2 \tablenotemark{g} & 0.003 (1) &   0.4 (1) & 790 (8)           & 100 (20) \\
HC$_3$N                    & $12-11$                     & 109.173634 & 34.1                   & 167.1 \tablenotemark{g} & $<0.005$  &  $< 0.5$  & \nodata           & \nodata  \\
OCS                        & $9-8$                       & 109.463063 & 26.3                   & 4.60                    & $<0.004$  &  $< 0.5$  & \nodata           & \nodata  \\
C$^{18}$O                  & $1-0$                       & 109.782173 &  5.3                   & 0.0122                  & 0.008 (5) &   1.7 (8) & 720 (30)          & 340 (90) \\
HNCO                       & $5_{0\,5}-4_{0\,4}$         & 109.905749 & 15.8                   & 12.5 \tablenotemark{g}  & $<0.005$  &  $< 0.5$  & \nodata           & \nodata  \\
$^{13}$CO                  & $1-0$                       & 110.201354 &  5.3                   & 0.0122                  & 0.042 (5) &  10.6 (9) & 728 (4)           & 220 (10) \\
C$^{17}$O                  & $1-0$                       & 112.359284 &  5.4                   & 0.0122                  & $<0.006$  &  $< 0.7$  & \nodata           & \nodata  \\
CN \tablenotemark{h}       & $N=1-0,J=1/2-1/2,F=1/2-3/2$ & 113.144157 &  5.4                   & 1.25                    & 0.013 (8) &   3 (1)   & \tablenotemark{i} & \nodata  \\
CN \tablenotemark{h}       & $N=1-0,J=1/2-1/2,F=3/2-1/2$ & 113.170492 &  5.4                   & 1.22                    &           &           & \nodata           & \nodata  \\
CN \tablenotemark{h}       & $N=1-0,J=1/2-1/2,F=3/2-3/2$ & 113.191279 &  5.4                   & 1.58                    &           &           & \nodata           & \nodata  \\
CN \tablenotemark{h}       & $N=1-0,J=3/2-1/2,F=5/2-3/2$ & 113.490970 &  5.4                   & 1.58                    & 0.018 (8) &   5 (1)   & \tablenotemark{i} & \nodata  \\
CN \tablenotemark{h}       & $N=1-0,J=3/2-1/2,F=3/2-1/2$ & 113.488120 &  5.4                   & 4.21                    &           &           & \nodata           & \nodata  \\
CN \tablenotemark{h}       & $N=1-0,J=3/2-1/2,F=1/2-1/2$ & 113.499644 &  5.4                   & 1.25                    &           &           & \nodata           & \nodata  \\
CO                         & $1-0$                       & 115.271202 &  5.5                   & 0.0121                  & 0.626 (7) & 144 (1)   & 737 (3)           & 219 (7)  \\
\tableline
\end{tabular}
\tablenotetext{a}{The numbers in parentheses represent $3\sigma$ errors.}
\tablenotetext{b}{Upper limits to the peak temperature are $3\sigma$.}
\tablenotetext{c}{The upper limit to the integrated intensity is calculated as: $\int T_{\rm mb} dv < 3 \sigma \times \sqrt{\Delta V \times \Delta v_{\rm res}}$, where $\Delta V$ is the assumed line width (220~km~s$^{-1}$) and $\Delta v_{\rm res}$ is the velocity resolution per channel.}
\tablenotetext{d}{The numbers in parentheses represent $1\sigma$ errors in units of the last significant digits.}
\tablenotetext{e}{The upper state energy is calculated from the lowest ortho state ($1_{0\,1}$).}
\tablenotetext{f}{The spin weight of 3 is not included.}
\tablenotetext{g}{The nuclear spin multiplicity of 3 for the N nucleus is not included.}
\tablenotetext{h}{The line is blended with other lines.}
\tablenotetext{i}{Gaussian fitting is not successful due to blending with other lines.}
\tablenotetext{j}{The upper state energy is calculated from the lowest E state ($1_{-1}$, E).}
\end{table}

\begin{table}
\tiny
\caption{Column Densities and Fractional Abundances in NGC~3627~SA \tablenotemark{ab} \label{tab08}}
\begin{tabular}{lllllll}
\tableline\tableline
Molecule       & $N$ \tablenotemark{c} ($T$ = 10~K \tablenotemark{d}) & $X$ \tablenotemark{e} ($T$ = 10~K \tablenotemark{d})          
               & $N$ \tablenotemark{c} ($T$ = 15~K \tablenotemark{d}) & $X$ \tablenotemark{e} ($T$ = 15~K \tablenotemark{d})          
               & $N$ \tablenotemark{c} ($T$ = 20~K \tablenotemark{d}) & $X$ \tablenotemark{e} ($T$ = 20~K \tablenotemark{d})  \\        
               & (cm$^{-2}$)                  &                           
               & (cm$^{-2}$)                  &                           
               & (cm$^{-2}$)                  &                   \\        
\tableline
CCH                              & $9.6 (3.5)  \times 10^{13}$  & $6.5 (2.4)  \times 10^{-9}$ 
                                 & $1.2 (0.4)  \times 10^{14}$  & $6.7 (2.5)  \times 10^{-9}$ 
                                 & $1.4 (0.5)  \times 10^{14}$  & $6.9 (2.5)  \times 10^{-9}$ \\
CN                               & $1.4 (0.5)  \times 10^{13}$  & $9.7 (3.2)  \times 10^{-10}$ 
                                 & $1.7 (0.6)  \times 10^{13}$  & $9.7 (3.2)  \times 10^{-10}$ 
                                 & $1.9 (0.7)  \times 10^{13}$  & $9.6 (3.2)  \times 10^{-10}$ \\
HCN                              & $6.5 (1.7)  \times 10^{12}$  & $4.4 (1.2)  \times 10^{-10}$ 
                                 & $7.7 (2.1)  \times 10^{12}$  & $4.5 (1.2)  \times 10^{-10}$ 
                                 & $9.2 (2.5)  \times 10^{12}$  & $4.6 (1.2)  \times 10^{-10}$ \\
H$^{13}$CN \tablenotemark{f}     & $< 5.4 \times 10^{11     }$  & $< 3.7 \times 10^{-11}$  
                                 & $< 6.5 \times 10^{11     }$  & $< 3.8 \times 10^{-11}$  
                                 & $< 7.8 \times 10^{11     }$  & $< 3.9 \times 10^{-11}$  \\
HNC                              & $1.9 (0.7)  \times 10^{12}$  & $1.3 (0.5)  \times 10^{-10}$ 
                                 & $2.3 (0.8)  \times 10^{12}$  & $1.3 (0.5)  \times 10^{-10}$ 
                                 & $2.7 (1.0)  \times 10^{12}$  & $1.3 (0.5)  \times 10^{-10}$ \\
HN$^{13}$C \tablenotemark{f}     & $< 5.0 \times 10^{11     }$  & $< 3.4 \times 10^{-11}$  
                                 & $< 6.1 \times 10^{11     }$  & $< 3.5 \times 10^{-11}$  
                                 & $< 7.2 \times 10^{11     }$  & $< 3.6 \times 10^{-11}$  \\
CO                               & $1.5 (0.3)  \times 10^{17}$  & $1.0 (0.2)  \times 10^{-5}$ 
                                 & $1.8 (0.4)  \times 10^{17}$  & $1.0 (0.2)  \times 10^{-5}$ 
                                 & $2.1 (0.4)  \times 10^{17}$  & $1.0 (0.2)  \times 10^{-5}$ \\
$^{13}$CO                        & $1.5 (0.3)  \times 10^{16}$  & $1.0 (0.2)  \times 10^{-6}$ 
                                 & $1.7 (0.4)  \times 10^{16}$  & $1.0 (0.2)  \times 10^{-6}$ 
                                 & $1.5 (0.3)  \times 10^{16}$  & $1.0 (0.2)  \times 10^{-6}$ \\
C$^{17}$O \tablenotemark{f}      & $< 2.2 \times 10^{14}$       & $< 1.5 \times 10^{-08}$  
                                 & $< 2.6 \times 10^{14}$       & $< 1.5 \times 10^{-08}$  
                                 & $< 3.0 \times 10^{14}$       & $< 1.5 \times 10^{-08}$  \\
C$^{18}$O                        & $2.5 (0.7)  \times 10^{15}$  &   
                                 & $2.9 (0.8)  \times 10^{15}$  &                     
                                 & $3.4 (0.9)  \times 10^{15}$  &                    \\   
HCO$^+$                          & $4.4 (1.0)  \times 10^{12}$  & $3.0 (0.7)  \times 10^{-10}$ 
                                 & $5.2 (1.2)  \times 10^{12}$  & $3.0 (0.7)  \times 10^{-10}$ 
                                 & $6.2 (1.5)  \times 10^{12}$  & $3.1 (0.7)  \times 10^{-10}$ \\
H$^{13}$CO$^+$ \tablenotemark{f} & $< 3.1 \times 10^{11}$       & $< 2.1 \times 10^{-11}$  
                                 & $< 3.8 \times 10^{11}$       & $< 2.2 \times 10^{-11}$  
                                 & $< 4.5 \times 10^{11}$       & $< 2.2 \times 10^{-11}$  \\
H$_2$CO \tablenotemark{g}        & $1.2 (0.8)  \times 10^{12}$  & $8.2 (5.2)  \times 10^{-11}$ 
                                 & $1.3 (0.8)  \times 10^{12}$  & $7.6 (4.8)  \times 10^{-11}$ 
                                 & $1.5 (0.9)  \times 10^{12}$  & $7.3 (4.6)  \times 10^{-11}$ \\
CH$_3$OH \tablenotemark{fh}      & $< 9.5 \times 10^{12}$       & $< 6.5 \times 10^{-10}$  
                                 & $< 1.3 \times 10^{13}$       & $< 7.6 \times 10^{-10}$  
                                 & $< 1.8 \times 10^{13}$       & $< 8.8 \times 10^{-10}$  \\ 
N$_2$H$^+$ \tablenotemark{f}     & $< 4.3 \times 10^{11}$       & $< 2.9 \times 10^{-11}$  
                                 & $< 5.1 \times 10^{11}$       & $< 3.0 \times 10^{-11}$  
                                 & $< 6.1 \times 10^{11}$       & $< 3.0 \times 10^{-11}$  \\
SiO \tablenotemark{f}            & $< 5.9 \times 10^{11}$       & $< 4.0 \times 10^{-11}$  
                                 & $< 6.7 \times 10^{11}$       & $< 3.9 \times 10^{-11}$  
                                 & $< 7.8 \times 10^{11}$       & $< 3.9 \times 10^{-11}$  \\
HNCO \tablenotemark{f}           & $< 4.4 \times 10^{12}$       & $< 3.0 \times 10^{-10}$  
                                 & $< 4.7 \times 10^{12}$       & $< 2.8 \times 10^{-10}$  
                                 & $< 5.7 \times 10^{13}$       & $< 2.8 \times 10^{-10}$  \\
c-C$_3$H$_2$ \tablenotemark{fg}  & $< 2.7 \times 10^{12}$       & $< 1.8 \times 10^{-10}$  
                                 & $< 3.7 \times 10^{12}$       & $< 2.2 \times 10^{-10}$  
                                 & $< 5.0 \times 10^{12}$       & $< 2.4 \times 10^{-10}$  \\
CH$_3$CCH \tablenotemark{fh}     & $< 4.7 \times 10^{13}$       & $< 3.2 \times 10^{-09}$  
                                 & $< 4.4 \times 10^{13}$       & $< 2.6 \times 10^{-09}$  
                                 & $< 4.9 \times 10^{13}$       & $< 2.4 \times 10^{-09}$  \\
CH$_3$CN \tablenotemark{fh}      & $< 1.2 \times 10^{12}$       & $< 7.9 \times 10^{-11}$  
                                 & $< 1.1 \times 10^{12}$       & $< 6.2 \times 10^{-11}$  
                                 & $< 1.2 \times 10^{12}$       & $< 5.7 \times 10^{-11}$  \\ 
CS                               & $4.7 (1.9)  \times 10^{12}$  & $3.2 (1.3)  \times 10^{-10}$ 
                                 & $5.2 (2.1)  \times 10^{12}$  & $3.0 (1.2)  \times 10^{-10}$ 
                                 & $6.0 (2.4)  \times 10^{12}$  & $3.0 (1.2)  \times 10^{-10}$ \\ 
C$^{34}$S \tablenotemark{f}      & $< 1.1 \times 10^{12}$       & $< 7.1 \times 10^{-11}$  
                                 & $< 1.2 \times 10^{12}$       & $< 6.8 \times 10^{-11}$  
                                 & $< 1.3 \times 10^{12}$       & $< 6.7 \times 10^{-11}$  \\
SO \tablenotemark{f}             & $< 2.6 \times 10^{12}$       & $< 1.8 \times 10^{-10}$  
                                 & $< 3.1 \times 10^{12}$       & $< 1.8 \times 10^{-10}$  
                                 & $< 3.7 \times 10^{12}$       & $< 1.8 \times 10^{-10}$  \\
HC$_3$N \tablenotemark{f}        & $< 2.2 \times 10^{12}$       & $< 1.5 \times 10^{-10}$  
                                 & $< 1.2 \times 10^{12}$       & $< 6.9 \times 10^{-11}$  
                                 & $< 9.5 \times 10^{11}$       & $< 4.7 \times 10^{-11}$  \\
OCS \tablenotemark{f}            & $< 4.0 \times 10^{13}$       & $< 2.7 \times 10^{-09}$  
                                 & $< 2.8 \times 10^{13}$       & $< 1.7 \times 10^{-09}$  
                                 & $< 2.6 \times 10^{13}$       & $< 1.3 \times 10^{-09}$  \\
\tableline
\end{tabular}
\tablenotetext{a}{The column densities and fractional abundances are derived by assuming the source size of $10''$.}
\tablenotetext{b}{Errors of the column densities are evaluated by taking into account of the rms noise ($3\sigma$) and the intensity calibration uncertainty of the chopper-wheel method (20~\%).  The numbers in parentheses represent the errors in units of the last significant digits.}
\tablenotetext{c}{Column density.}
\tablenotetext{d}{Assumed excitation temperature.}
\tablenotetext{e}{Fractional abundance relative to the H$_2$.  The column density of H$_2$ is derived from the column density of C$^{18}$O, where $N$(C$^{18}$O)/$N$(H$_2$) of $1.7 \times 10^{-7}$ is assumed \citep{Frerking1982,Goldsmith1997}.  The errors of the fractional abundances do not include the error of H$_2$ column density.}
\tablenotetext{f}{The upper limit is derived from the $3\sigma$ upper limit of the integrated intensity.}
\tablenotetext{g}{The column density is calculated from the ortho species by assuming the ortho-to-para ratio of 3.}
\tablenotetext{h}{The column density is calculated from the A species ($K = 0$) on the assumption that the column density of the E species is the same as that of the A species.}
\end{table}

\begin{table}
\tiny
\caption{Column Densities and Fractional Abundances in NGC~3627~BE \tablenotemark{ab} \label{tab09}}
\begin{tabular}{lllllll}
\tableline\tableline
Molecule       & $N$ \tablenotemark{c} ($T$ = 10~K \tablenotemark{d}) & $X$ \tablenotemark{e} ($T$ = 10~K \tablenotemark{d})          
               & $N$ \tablenotemark{c} ($T$ = 15~K \tablenotemark{d}) & $X$ \tablenotemark{e} ($T$ = 15~K \tablenotemark{d})          
               & $N$ \tablenotemark{c} ($T$ = 20~K \tablenotemark{d}) & $X$ \tablenotemark{e} ($T$ = 20~K \tablenotemark{d})  \\        
               & (cm$^{-2}$)                  &                           
               & (cm$^{-2}$)                  &                           
               & (cm$^{-2}$)                  &                   \\        
\tableline
CCH                              & $2.3 (0.7)  \times 10^{14}$  & $7.6 (2.4)  \times 10^{-09}$ 
                                 & $2.8 (0.9)  \times 10^{14}$  & $7.9 (2.4)  \times 10^{-09}$
                                 & $3.3 (1.0)  \times 10^{14}$  & $8.0 (2.5)  \times 10^{-09}$ \\ 
CN                               & $6.3 (1.4)  \times 10^{13}$  & $2.1 (0.5)  \times 10^{-09}$ 
                                 & $7.3 (1.6)  \times 10^{13}$  & $2.1 (0.5)  \times 10^{-09}$
                                 & $8.5 (1.9)  \times 10^{13}$  & $2.1 (0.5)  \times 10^{-09}$ \\ 
HCN                              & $2.7 (0.6)  \times 10^{13}$  & $9.0 (2.0)  \times 10^{-10}$ 
                                 & $3.2 (0.7)  \times 10^{13}$  & $9.3 (2.1)  \times 10^{-10}$
                                 & $3.9 (0.9)  \times 10^{13}$  & $9.4 (2.1)  \times 10^{-10}$ \\ 
H$^{13}$CN \tablenotemark{f}     & $< 1.6 \times 10^{12}$       & $< 5.4 \times 10^{-11}$ 
                                 & $< 2.0 \times 10^{12}$       & $< 5.6 \times 10^{-11}$
                                 & $< 2.3 \times 10^{12}$       & $< 5.7 \times 10^{-11}$ \\ 
HNC                              & $8.8 (2.7)  \times 10^{12}$  & $2.9 (0.9)  \times 10^{-10}$ 
                                 & $1.1 (0.3)  \times 10^{13}$  & $3.0 (0.9)  \times 10^{-10}$
                                 & $1.3 (0.4)  \times 10^{13}$  & $3.0 (0.9)  \times 10^{-10}$ \\ 
HN$^{13}$C \tablenotemark{f}     & $< 1.9 \times 10^{12}$       & $< 6.1 \times 10^{-11}$ 
                                 & $< 2.2 \times 10^{12}$       & $< 6.3 \times 10^{-11}$
                                 & $< 2.7 \times 10^{12}$       & $< 6.4 \times 10^{-11}$ \\ 
CO                               & $2.7 (0.6)  \times 10^{17}$  & $9.0 (1.8)  \times 10^{-06}$ 
                                 & $3.1 (0.6)  \times 10^{17}$  & $9.0 (1.8)  \times 10^{-06}$
                                 & $3.7 (0.7)  \times 10^{17}$  & $8.9 (1.8)  \times 10^{-06}$ \\ 
$^{13}$CO                        & $2.9 (0.6)  \times 10^{16}$  & $9.6 (1.9)  \times 10^{-07}$ 
                                 & $3.4 (0.7)  \times 10^{16}$  & $9.6 (1.9)  \times 10^{-07}$
                                 & $4.0 (0.8)  \times 10^{16}$  & $9.6 (1.9)  \times 10^{-07}$ \\ 
C$^{17}$O                        & $7.2 (6.3)  \times 10^{14}$  & $2.4 (2.1)  \times 10^{-08}$ 
                                 & $8.4 (7.3)  \times 10^{14}$  & $2.4 (2.1)  \times 10^{-08}$
                                 & $9.8 (8.6)  \times 10^{14}$  & $2.4 (2.1)  \times 10^{-08}$ \\  
C$^{18}$O                        & $5.1 (1.2)  \times 10^{15}$  &                    
                                 & $6.0 (1.4)  \times 10^{15}$  &                   
                                 & $7.0 (1.6)  \times 10^{15}$  &                    \\ 
HCO$^+$                          & $1.4 (0.3)  \times 10^{13}$  & $4.7 (1.0)  \times 10^{-10}$ 
                                 & $1.7 (0.5)  \times 10^{13}$  & $4.8 (1.1)  \times 10^{-10}$
                                 & $2.0 (0.5)  \times 10^{13}$  & $4.9 (1.1)  \times 10^{-10}$ \\ 
H$^{13}$CO$^+$ \tablenotemark{f} & $< 1.0 \times 10^{12}$       & $< 3.5 \times 10^{-11}$ 
                                 & $< 1.3 \times 10^{12}$       & $< 3.6 \times 10^{-11}$
                                 & $< 1.5 \times 10^{12}$       & $< 3.6 \times 10^{-11}$ \\
H$_2$CO \tablenotemark{g}        & $6.9 (1.9)  \times 10^{12}$  & $2.3 (0.6)  \times 10^{-10}$ 
                                 & $7.4 (2.1)  \times 10^{12}$  & $2.1 (0.6)  \times 10^{-10}$
                                 & $8.4 (2.4)  \times 10^{12}$  & $2.0 (0.6)  \times 10^{-10}$ \\ 
CH$_3$OH \tablenotemark{fh}      & $6.7 (2.3)  \times 10^{13}$  & $2.2 (0.8)  \times 10^{-09}$ 
                                 & $9.2 (3.2)  \times 10^{13}$  & $2.6 (0.9)  \times 10^{-09}$
                                 & $1.2 (0.4)  \times 10^{14}$  & $3.0 (1.0)  \times 10^{-09}$ \\
N$_2$H$^+$                       & $2.3 (1.0)  \times 10^{12}$  & $7.5 (3.2)  \times 10^{-11}$ 
                                 & $2.7 (1.2)  \times 10^{12}$  & $7.7 (3.3)  \times 10^{-11}$
                                 & $3.2 (1.4)  \times 10^{12}$  & $7.8 (3.3)  \times 10^{-11}$ \\
SiO \tablenotemark{f}            & $< 2.2 \times 10^{12}$       & $< 7.2 \times 10^{-11}$ 
                                 & $< 2.5 \times 10^{12}$       & $< 7.0 \times 10^{-11}$
                                 & $< 2.9 \times 10^{12}$       & $< 6.9 \times 10^{-11}$ \\
HNCO \tablenotemark{f}           & $< 1.2 \times 10^{13}$       & $< 4.0 \times 10^{-10}$ 
                                 & $< 1.3 \times 10^{13}$       & $< 3.7 \times 10^{-10}$
                                 & $< 1.6 \times 10^{13}$       & $< 3.8 \times 10^{-10}$ \\
c-C$_3$H$_2$ \tablenotemark{fi}  & $< 6.0 \times 10^{12}$       & $< 2.0 \times 10^{-10}$ 
                                 & $< 8.3 \times 10^{12}$       & $< 2.4 \times 10^{-10}$
                                 & $< 1.1 \times 10^{13}$       & $< 2.7 \times 10^{-10}$ \\
CH$_3$CCH \tablenotemark{fh}     & $< 1.3 \times 10^{14}$       & $< 4.3 \times 10^{-09}$ 
                                 & $< 1.2 \times 10^{14}$       & $< 3.5 \times 10^{-09}$
                                 & $< 1.3 \times 10^{14}$       & $< 3.2 \times 10^{-09}$ \\
CH$_3$CN \tablenotemark{fh}      & $< 4.2 \times 10^{12}$       & $< 1.4 \times 10^{-10}$ 
                                 & $< 3.9 \times 10^{12}$       & $< 1.1 \times 10^{-10}$
                                 & $< 4.2 \times 10^{12}$       & $< 1.0 \times 10^{-10}$ \\
CS                               & $1.7 (0.5)  \times 10^{13}$  & $5.8 (1.5)  \times 10^{-10}$ 
                                 & $1.9 (0.5)  \times 10^{13}$  & $5.5 (1.5)  \times 10^{-10}$
                                 & $2.2 (0.6)  \times 10^{13}$  & $5.4 (1.4)  \times 10^{-10}$ \\
C$^{34}$S \tablenotemark{f}      & $4.6 (3.3)  \times 10^{12}$  & $1.5 (1.1)  \times 10^{-10}$ 
                                 & $5.1 (3.7)  \times 10^{12}$  & $1.4 (1.0)  \times 10^{-10}$
                                 & $5.8 (4.2)  \times 10^{12}$  & $1.4 (1.0)  \times 10^{-10}$ \\
SO                               & $1.1 (0.7)  \times 10^{13}$  & $3.7 (2.4)  \times 10^{-10}$ 
                                 & $1.3 (0.9)  \times 10^{13}$  & $3.8 (2.4)  \times 10^{-10}$
                                 & $1.6 (1.0)  \times 10^{13}$  & $3.9 (2.5)  \times 10^{-10}$ \\
HC$_3$N \tablenotemark{f}        & $< 3.6 \times 10^{12}$       & $< 1.2 \times 10^{-10}$ 
                                 & $< 2.0 \times 10^{12}$       & $< 5.6 \times 10^{-11}$
                                 & $< 1.6 \times 10^{12}$       & $< 3.8 \times 10^{-11}$ \\
OCS \tablenotemark{f}            & $< 5.0 \times 10^{13}$       & $< 1.7 \times 10^{-09}$ 
                                 & $< 3.5 \times 10^{13}$       & $< 1.0 \times 10^{-09}$
                                 & $< 3.2 \times 10^{13}$       & $< 7.8 \times 10^{-10}$ \\
\tableline
\end{tabular}
\tablenotetext{a}{The column densities and fractional abundances are derived by assuming the source size of $10''$.}
\tablenotetext{b}{Errors of the column densities are evaluated by taking into account of the rms noise ($3\sigma$) and the intensity calibration uncertainty of the chopper-wheel method (20~\%).  The numbers in parentheses represent the errors in units of the last significant digits.}
\tablenotetext{c}{Column density.}
\tablenotetext{d}{Assumed excitation temperature.}
\tablenotetext{e}{Fractional abundance relative to the H$_2$.  The column density of H$_2$ is derived from the column density of C$^{18}$O, where $N$(C$^{18}$O)/$N$(H$_2$) of $1.7 \times 10^{-7}$ is assumed \citep{Frerking1982,Goldsmith1997}.  The errors of the fractional abundances do not include the error of H$_2$ column density.}
\tablenotetext{f}{The upper limit is derived from the $3\sigma$ upper limit of the integrated intensity.}
\tablenotetext{g}{The total column density of the ortho species and the para species.}
\tablenotetext{h}{The column density is calculated from the A species ($K = 0$) on the assumption that the column density of the E species is the same as that of the A species.}
\tablenotetext{i}{The column density is calculated from the ortho species by assuming the ortho-to-para ratio of 3.}
\end{table}

\begin{table}
\tiny
\caption{Column Densities and Fractional Abundances in NGC~3627~NR \tablenotemark{ab} \label{tab10}}
\begin{tabular}{lllllll}
\tableline\tableline
Molecule       & $N$ \tablenotemark{c} ($T$ = 10~K \tablenotemark{d}) & $X$ \tablenotemark{e} ($T$ = 10~K \tablenotemark{d})          
               & $N$ \tablenotemark{c} ($T$ = 15~K \tablenotemark{d}) & $X$ \tablenotemark{e} ($T$ = 15~K \tablenotemark{d})          
               & $N$ \tablenotemark{c} ($T$ = 20~K \tablenotemark{d}) & $X$ \tablenotemark{e} ($T$ = 20~K \tablenotemark{d})  \\        
               & (cm$^{-2}$)                  &                           
               & (cm$^{-2}$)                  &                           
               & (cm$^{-2}$)                  &                   \\        
\tableline
CCH                              & $1.9 (1.0) \times 10^{14}$  & $5.5 (2.9) \times 10^{-9}$ 
                                 & $2.3 (1.2) \times 10^{14}$  & $5.7 (3.0) \times 10^{-9}$ 
                                 & $2.7 (1.4) \times 10^{14}$  & $5.8 (3.0) \times 10^{-9}$ \\
CN                               & $1.3 (4.4) \times 10^{14}$  & $3.6 (1.3) \times 10^{-9}$ 
                                 & $1.5 (5.1) \times 10^{14}$  & $3.6 (1.3) \times 10^{-9}$ 
                                 & $1.7 (6.0) \times 10^{14}$  & $3.6 (1.3) \times 10^{-9}$ \\
HCN                              & $5.3 (1.3) \times 10^{13}$  & $1.5 (0.4) \times 10^{-9}$ 
                                 & $6.4 (1.5) \times 10^{13}$  & $1.6 (0.4) \times 10^{-9}$ 
                                 & $7.6 (1.8) \times 10^{13}$  & $1.6 (0.4) \times 10^{-9}$ \\
H$^{13}$CN \tablenotemark{f}     & $< 7.3 \times 10^{12}$      & $< 2.1 \times 10^{-10}$   
                                 & $< 8.8 \times 10^{12}$      & $< 2.2 \times 10^{-10}$   
                                 & $< 1.1 \times 10^{13}$      & $< 2.2 \times 10^{-10}$ \\
HNC                              & $1.4 (0.5) \times 10^{13}$  & $4.1 (1.4) \times 10^{-10}$ 
                                 & $1.7 (0.6) \times 10^{13}$  & $4.2 (1.4) \times 10^{-10}$ 
                                 & $2.0 (0.7) \times 10^{13}$  & $4.3 (1.4) \times 10^{-10}$ \\
HN$^{13}$C \tablenotemark{f}     & $< 2.8 \times 10^{12}$      & $< 8.2 \times 10^{-11}$   
                                 & $< 3.4 \times 10^{12}$      & $< 8.4 \times 10^{-11}$   
                                 & $< 4.0 \times 10^{12}$      & $< 8.6 \times 10^{-11}$ \\
CO                               & $4.3 (0.9) \times 10^{17}$  & $1.2 (0.3) \times 10^{-5}$ 
                                 & $4.9 (1.0) \times 10^{17}$  & $1.2 (0.3) \times 10^{-5}$ 
                                 & $5.8 (1.2) \times 10^{17}$  & $1.2 (0.2) \times 10^{-5}$ \\
$^{13}$CO                        & $3.5 (0.8) \times 10^{16}$  & $1.0 (0.2) \times 10^{-6}$ 
                                 & $4.1 (0.9) \times 10^{16}$  & $1.0 (0.2) \times 10^{-6}$ 
                                 & $4.8 (1.1) \times 10^{16}$  & $1.0 (0.2) \times 10^{-6}$ \\
C$^{17}$O \tablenotemark{f}      & $< 2.1 \times 10^{15}$      & $< 6.1 \times 10^{-8}$     
                                 & $< 2.4 \times 10^{15}$      & $< 6.1 \times 10^{-8}$   
                                 & $< 2.9 \times 10^{15}$      & $< 6.0 \times 10^{-8}$ \\
C$^{18}$O                        & $5.9 (3.0) \times 10^{15}$  &                 
                                 & $8.0 (4.1) \times 10^{15}$  &                 
                                 & $6.8 (3.5) \times 10^{15}$  &                             \\
HCO$^+$                          & $2.4 (0.6) \times 10^{13}$  & $6.9 (1.6) \times 10^{-10}$ 
                                 & $2.8 (0.7) \times 10^{13}$  & $7.1 (1.7) \times 10^{-10}$ 
                                 & $3.4 (0.8) \times 10^{13}$  & $7.2 (1.7) \times 10^{-10}$ \\
H$^{13}$CO$^+$ \tablenotemark{f} & $< 2.6 \times 10^{12}$      & $< 7.7 \times 10^{-11}$   
                                 & $< 3.2 \times 10^{12}$      & $< 7.9 \times 10^{-11}$   
                                 & $< 3.8 \times 10^{12}$      & $< 8.0 \times 10^{-11}$ \\
CH$_3$OH \tablenotemark{g}       & $1.7 (0.9) \times 10^{14}$  & $4.9 (2.5) \times 10^{-9}$ 
                                 & $2.3 (1.2) \times 10^{14}$  & $5.8 (3.0) \times 10^{-9}$ 
                                 & $3.1 (1.6) \times 10^{14}$  & $6.7 (3.4) \times 10^{-9}$ \\
N$_2$H$^+$ \tablenotemark{f}     & $< 2.0 \times 10^{12}$      & $< 5.8 \times 10^{-11}$   
                                 & $< 2.4 \times 10^{12}$      & $< 5.9 \times 10^{-11}$   
                                 & $< 2.8 \times 10^{12}$      & $< 6.0 \times 10^{-11}$ \\
SiO \tablenotemark{f}            & $< 4.1 \times 10^{12}$      & $< 1.2 \times 10^{-10}$   
                                 & $< 4.7 \times 10^{12}$      & $< 1.2 \times 10^{-10}$   
                                 & $< 5.5 \times 10^{12}$      & $< 1.2 \times 10^{-10}$ \\
HNCO \tablenotemark{f}           & $< 1.9 \times 10^{13}$      & $< 5.6 \times 10^{-10}$   
                                 & $< 2.1 \times 10^{13}$      & $< 5.1 \times 10^{-10}$   
                                 & $< 2.5 \times 10^{13}$      & $< 5.2 \times 10^{-10}$ \\
c-C$_3$H$_2$ \tablenotemark{fh}  & $< 2.6 \times 10^{13}$      & $< 7.5 \times 10^{-10}$   
                                 & $< 3.6 \times 10^{13}$      & $< 8.9 \times 10^{-10}$   
                                 & $< 4.8 \times 10^{13}$      & $< 1.0 \times 10^{-9}$ \\
CH$_3$CCH \tablenotemark{fg}     & $< 4.5 \times 10^{14}$      & $< 1.3 \times 10^{-8}$   
                                 & $< 4.2 \times 10^{14}$      & $< 1.1 \times 10^{-8}$   
                                 & $< 4.7 \times 10^{14}$      & $< 1.0 \times 10^{-8}$ \\
CH$_3$CN \tablenotemark{fg}      & $< 8.0 \times 10^{12}$      & $< 2.3 \times 10^{-10}$   
                                 & $< 7.3 \times 10^{12}$      & $< 1.8 \times 10^{-10}$   
                                 & $< 8.0 \times 10^{12}$      & $< 1.7 \times 10^{-10}$ \\
CS                               & $5.4 (1.2) \times 10^{13}$  & $1.6 (0.3) \times 10^{-9}$ 
                                 & $6.0 (1.3) \times 10^{13}$  & $1.5 (0.3) \times 10^{-9}$ 
                                 & $6.9 (1.5) \times 10^{13}$  & $1.5 (0.3) \times 10^{-9}$ \\
C$^{34}$S \tablenotemark{f}      & $< 9.5 \times 10^{12}$      & $< 2.7 \times 10^{-10}$   
                                 & $< 1.1 \times 10^{13}$      & $< 2.6 \times 10^{-10}$   
                                 & $< 1.2 \times 10^{13}$      & $< 2.6 \times 10^{-10}$ \\
SO \tablenotemark{f}             & $< 2.2 \times 10^{13}$      & $< 6.4 \times 10^{-10}$   
                                 & $< 2.6 \times 10^{13}$      & $< 6.5 \times 10^{-10}$   
                                 & $< 3.1 \times 10^{13}$      & $< 6.7 \times 10^{-10}$ \\
HC$_3$N                          & $1.8 (0.6) \times 10^{13}$  & $5.1 (1.8) \times 10^{-10}$ 
                                 & $9.6 (3.4) \times 10^{12}$  & $2.4 (0.9) \times 10^{-10}$ 
                                 & $7.7 (2.8) \times 10^{12}$  & $1.6 (0.6) \times 10^{-10}$ \\
OCS \tablenotemark{f}            & $< 1.4 \times 10^{14}$      & $< 4.2 \times 10^{-09}$   
                                 & $< 1.0 \times 10^{14}$      & $< 2.5 \times 10^{-09}$   
                                 & $< 9.3 \times 10^{13}$      & $< 2.0 \times 10^{-09}$ \\ 
\tableline
\end{tabular}
\tablenotetext{a}{The column densities and fractional abundances are derived by assuming the source size of $10''$.}
\tablenotetext{b}{Errors of the column densities are evaluated by taking into account of the rms noise ($3\sigma$) and the intensity calibration uncertainty of the chopper-wheel method (20~\%).  The numbers in parentheses represent the errors in units of the last significant digits.}
\tablenotetext{c}{Column density.}
\tablenotetext{d}{Assumed excitation temperature.}
\tablenotetext{e}{Fractional abundance relative to the H$_2$.  The column density of H$_2$ is derived from the column density of C$^{18}$O, where $N$(C$^{18}$O)/$N$(H$_2$) of $1.7 \times 10^{-7}$ is assumed \citep{Frerking1982,Goldsmith1997}.  The errors of the fractional abundances do not include the error of H$_2$ column density.}
\tablenotetext{f}{The upper limit is derived from the $3\sigma$ upper limit of the integrated intensity.}
\tablenotetext{g}{The column density is calculated from the A species ($K = 0$) on the assumption that the column density of the E species is the same as that of the A species.}
\tablenotetext{h}{The column density is calculated from the ortho species by assuming the ortho-to-para ratio of 3.}
\end{table}

\end{document}